%% file: main.tex
\documentclass[pdftex,twocolumn,epjc3,envcountsect]{svjour3}          % twocolumn

\smartqed
\usepackage{caption}
\usepackage[freestanding]{hepunits}
\usepackage{subcaption}
\usepackage{xparse}
\usepackage{siunitx}[=v2]

\RequirePackage{fix-cm}
\RequirePackage{graphicx}
\RequirePackage{grffile}
\RequirePackage{mathptmx}      % use Times fonts if available on your TeX system
\RequirePackage{flushend}
\RequirePackage[colorlinks,citecolor=blue,urlcolor=blue,linkcolor=blue]{hyperref}
\RequirePackage{amsmath}
\RequirePackage{cite}
\RequirePackage{multirow}
\RequirePackage[modulo]{lineno}

\graphicspath{{figs/}}

\journalname{Eur. Phys. J. C}

\newcommand{\mailto}[1]{\href{mailto:#1}{\texttt{#1}}}

\newcommand{\theia}{\textsc{Theia}}
\newcommand{\eos}{\textsc{Eos}}
\newcommand{\wbls}{WbLS}

\newcommand{\isotope}[2]{$^{#2}$#1}
\newcommand{\tz}{t_{0}}
\newcommand{\pp}[1]{\left(#1\right)}
\newcommand{\tR}{\tau_{R}}
\newcommand{\eto}[1]{e^{#1}}
\newcommand{\ndf}{\text{ndf}}
\newcommand{\differential}[1]{\text{d}#1}
\newcommand{\plusminus}[2]{^{+#1}_{-#2}}

\newcommand{\derivative}[2]{\frac{\differential{#1}}{\differential{#2}}}
\newcommand{\dLdx}{\derivative{L}{x}}
\newcommand{\dEdx}{\derivative{E}{x}}
\newcommand{\kB}{kB}
\newcommand{\inch}{in}
\newcommand{\ccut}{c'}

\newcommand{\pid}{PID}
\newcommand{\jax}{\texttt{JAX}}

\newcommand{\geant}{\texttt{GEANT4}}
\newcommand{\ratpac}{\texttt{RAT-PAC}}
\newcommand{\glgfscint}{\texttt{GLG4Scint}}

\begin{document} \sloppy % Fixes the line-break issues

\title{Characterization of the scintillation response of water-based liquid scintillator to alpha particles, and implications for particle identification}

\author{E. J. Callaghan \thanksref{ucb,lbl,e1}
\and
T. Kaptanoglu \thanksref{ucb,lbl,e2}
\and
M. Smiley \thanksref{ucb,lbl,e3}
\and
M. Yeh \thanksref{bnl}
\and
G. D. Orebi Gann\thanksref{ucb,lbl}
}

\institute{University of California, Berkeley, CA 94720-7300, USA \label{ucb}
           \and
           Lawrence Berkeley National Laboratory, CA 94720-8153, USA \label{lbl}
           \and
	   Brookhaven National Laboratory, Upton, NY 11973-500, USA \label{bnl}
}

\thankstext{e1}{email: \mailto{ejc3@berkeley.edu}}
\thankstext{e2}{email: \mailto{tannerbk@berkeley.edu}}
\thankstext{e3}{email: \mailto{masmiley@berkeley.edu}}

\date{Received: date / Accepted: date}

\maketitle

\begin{abstract}

    Next-generation large-scale neutrino detectors, from \eos{}, at the \SI{1}{\tonne{}} scale, to \theia{}, at the 10s-of-\SI{}{\kilo\tonne{}} scale, will utilize differences in both the scintillation and Cherenkov light emission for different particle species to perform background rejection. This manuscript presents measurements of the scintillation light yield and emission time profile of water-based liquid scintillator samples in response to $\alpha$ radiation. These measurements are used as input to simulation models used to make predictions for future detectors. In particular, we present the timing-based particle identification achievable in generic water-based scintillator detectors at the \SI{4}{\tonne{}}, \SI{1}{\kilo\tonne{}}, and \SI{100}{\kilo\tonne{}} scales. We find that $\alpha$/$\beta$ discrimination improves with increasing scintillation concentration and we identify better than 80\% $\alpha$ rejection for 90\% $\beta$ acceptance in 10\% water-based liquid scintillator, at the \SI{4}{\tonne{}} scale.

\end{abstract}

\input{intro}

\input{simulations}

\input{ly}

\input{timing}

\input{pid}

\input{conclusion}

\input{acknowledgements}

%%%% Bibliography

\clearpage

\bibliographystyle{ieeetr}
\bibliography{main.bib}

\end{document}

%% file: intro.tex
\section{\label{sec:intro} Introduction}

    Large liquid-phase optical detectors have seen continued use in neutrino physics, and are responsible for a number of key results, ranging from the discovery of neutrino flavor transformation to sensitive probes of the neutrino oscillation patterns \cite{Ahmad:2002jz,Agostini:2021bxc,Ahn:2006zza,IMB,kamland,Adamson:2017gxd,Abe:2019vii,Seo:2019shs,MiniBooNe:2013uba}. Historically, water detectors utilized Cherenkov light, whereas detectors employing a liquid scintillator overwhelmingly made use of scintillation light alone. A hybrid detector, which would simultaneously leverage both Cherenkov and scintillation signals for advanced event reconstruction, is a strong candidate for a next-generation detector. One design is \theia{} \cite{Askins:2019oqj}, a proposed tens of kilotonnes detector which would support a broad experimental program. Fundamental physics topics include measurements of low-energy solar neutrinos \cite{Bonventre:2018hyd,Land:2020oiz} and reactor- and geo-antineutrinos \cite{Zsoldos:2022mre}, searches for neutrinoless double-beta decay \cite{Askins:2019oqj,Land:2020oiz}, determination of the CP-violating phase of the PMNS matrix, and neutrino mass heirarchy \cite{Askins:2019oqj,snomass_theia_loi}. As a large-scale antineutrino detector, \theia{} could act as a proof-of-principle for far-field reactor monitoring, in the context of nuclear nonproliferation \cite{WATCHMAN:2015lcq,NuTools,ArmsControl}.

    A challenge in addressing many of these topics experimentally is the rejection of radioactive backgrounds, typically in the low-energy regime around \SI{1}{~\MeV}. One such class of background is associated with $\alpha$ radiation from the series of decay products in the uranium and thorium decay series. This radiation can manifest as a background to measurements of elastic scattering of low-energy solar neutrinos \cite{Borexino:2013zhu}. Additionally \isotope{Bi}{212}-\isotope{Po}{212} and \isotope{Bi}{214}-\isotope{Po}{214} decays, as well as generic $\pp{\alpha,n}$ reactions on target nuclei \cite{Zhao:2013mba}, constitute time-correlated events, which can mimic inverse-beta decay events used to detect antineutrinos \cite{Zsoldos:2022mre,SNO:2015wyx}.

    In liquid scintillators, these events can be classified as $\alpha$-induced using timing-based particle-identification (\pid{}), which exploits the different scintillation emission time profiles exhibited by particles with different ionization characteristics \cite{Ranucci:1994cs}. This is often achieved with a method known as pulse shape discrimination (PSD), which leverages ratios of the amounts of light observed over different time windows, often computed from a single multiphoton waveform, to classify events as $\alpha$-like or $\beta$-like. This can be extended further by considering the time of each detected photon individually, and using a likelihood-ratio test to compare the $\alpha$ and $\beta$ hypotheses.

    Such timing-based PID has been utilized in past liquid scintillator detectors, for example Borexino, which demonstrated low-energy $\alpha$/$\beta$ discrimination on the basis of scintillation timing using a Gatti filter PSD method \cite{borexinoab}, and in SNO+, which aims to use a likelihood-ratio-based discriminant \cite{SNO:2020fhu}. In a hybrid detector capable of identifying Cherenkov light, additional PID is possible beyond that provided by differences in scintillation emission timing, as heavy ions, including $\alpha$s and protons, are below the Cherenkov-threshold at electron-equivalent energies below the $\sim \SI{100}{~\MeV}$ scale. Using only timing information, the additional discrimination power available will be limited at low energy, where there are relatively few Cherenkov photons compared to the scintillation light. But future hybrid detectors, which will employ sophisticated Cherenkov tagging, can use the ratio of the number of detected Cherenkov and scintillation photons as a PID metric, for example to reject hadronic events from atmospheric neutral current reactions, which form a background to reactor- and geo-neutrino analyses, as well as to searches for the diffuse supernova neutrino background \cite{Askins:2019oqj,Sawatzki:2020mpb}.

    A powerful tool for next-generation hybrid detectors is the separation of the scintillation and Cherenkov components of the observed photons, which can be achieved via fast-timing photosensors \cite{Kaptanoglu:2021prv,Caravaca:2016fjg}, spectral sorting of photons \cite{Kaptanoglu:2019gtg,Kaptanoglu:2018sus}, and sophisticated reconstruction algorithms \cite{Anderson:2022lbb}. At the materials level, the development of water-based liquid scintillator (\wbls{}) \cite{YehWbLS} has constituted a step toward hybrid detection technology in producing a scintillator with an intrinsically favorable proportion of Cherenkov light. \wbls{} is formed as a homogeneous mixture of linear alkylbenzene (LAB), 2,5-diphenyloxazole (PPO), and water. LAB and PPO are a common solvent-fluor pair which constitute a two-component system previously used in experimental neutrino physics \cite{SNO:2020fhu,DayaBay:2015kir}. There are further advantages of \wbls{} over pure liquid scintillator, including its relatively low cost, being water-based, and better environmental compatibility. Notably, \wbls{} is a flexible material in that the scintillator fraction can be tuned to adjust the light yield to suit detector goals. The scintillator loading in \wbls{} is variable over a broad range. In this work we perform measurements using alpha and electron sources with 1\%, 5\% and 10\% scintillator concentrations in the \wbls{} targets. The light yield and emission time profile of \wbls{} in response to electrons have been previously reported \cite{Caravaca:2020lfs,Kaptanoglu:2021prv}.

    Deployment of \wbls{} in demonstration-scale detectors is underway, with further usage planned. The ANNIE detector has deployed \SI{365}{\kilo\gram} of \wbls{} in a contained vessel to study high-energy neutrino reconstruction \cite{sandi}; \SI{1}{\tonne{}} has been deployed at Brookhaven National Laboratory to demonstrate long-term chemical stability and recirculation techniques, with a planned scale-up to \SI{30}{\tonne{}} underway; \eos{} will deploy \SI{4}{\tonne{}} of \wbls{} in a fiducialized detector to demonstrate low-energy direction reconstruction \cite{Anderson:2022lbb}; BUTTON \cite{button} will be an underground testbed that allows for studies of low-background capabilities; and RNET \cite{Bat:2021jyq} is a proposed program to demonstrate reactor antineutrino detection using a \wbls{} target as part of a broad reactor monitoring program for nuclear non-proliferation goals.

    This work presents a characterization of the light yield and time profile of \wbls{} in response to $\alpha$ radiation, and implications for timing-based PID in realistic detector deployments. Section~\ref{sec:simulations} introduces the simulation software used to measure the light yield and investigate the PID performance of realistic detectors. Section~\ref{sec:light_yield} and Sect.~\ref{sec:timing} describe benchtop measurements of the light yield and emission time profile in response to $\alpha$s, respectively. The results of these measurements are used as input into simulations of realistic detectors at the medium- and large-scale, and Sect.~\ref{sec:pid} presents the timing-based \pid{} performance achievable in such detectors utilizing both \wbls{} and pure LAB with PPO loaded at \SI{2}{~\gram/\liter} (LAB+PPO). Sect.~\ref{sec:conclusion} provides concluding thoughts and avenues for continued investigation.

%% file: simulations.tex
\section{\label{sec:simulations} Simulations}

    A detailed Monte Carlo (MC) simulation is implemented in \ratpac{} \cite{ratpac}, a \geant-based \cite{GEANT4:2002zbu} simulation package that models the interactions of radiation with matter, as well the production, propagation, and detection of photons. \ratpac{} allows for flexible detector geometry design, which is utilized in Sect.~\ref{sec:light_yield} to model the light yield experimental setup and in Sect.~\ref{sec:pid} to study PID in several large-scale detectors filled with scintillator.

    The production of the scintillation photons is performed using a \glgfscint{}-based optical model \cite{ratpac}, which further accounts for absorption, reemission, and Rayleigh scattering of the photons as they propagate through the detector medium. The inputs to the \wbls{} optical model are either measured \cite{Caravaca:2020lfs,Onken:2020pnv,Kaptanoglu:2021prv} or estimated using the component properties of water and LAB-based scintillator \cite{Caravaca:2020lfs}.

    The photomultiplier tubes (PMTs) are modeled as full 3D objects with geometries and quantum efficiencies taken from the manufacturer specifications \cite{ham_datasheet_large_area, r14688}. The PMTs are assigned properties, such as dark-rates and transit times, based on benchtop measurements specific to each PMT model. The methods for determining the properties important for the measurements presented in this manuscript, such as the single-photoelectron (SPE) pulse characteristics, are described in Sect.~\ref{sec:light_yield_calibrations}.

%% file: ly.tex
\section{Light Yield}\label{sec:light_yield}

    The scintillation light yield of three target \wbls{} samples, at nominal scintillator loadings of 1\%, 5\% and 10\%, under both $\beta$ and $\alpha$ excitation is measured. Previous light yield measurementsof \wbls{} under $\beta$ excitation were reported in Ref. \cite{Caravaca:2020lfs}. Light yield measurements are not performed for LAB+PPO, as the total amount of light saturates the dynamic range of the electronics, which are configured for sensitivity to samples with lower light levels, and would require dedicated recalibration.

    Ionization quenching refers to a reduction in scintillation output associated with local ionization density, which manifests as a nonlinear relation between scintillation light emitted and energy deposited in the material. A first semi-empirical model due to Birks \cite{birks}, which uses the material stopping power as a proxy for the ionization density, remains in use today. Birks' law relates the rate of production of scintillation photons, $L$, to the material stopping power as
\begin{equation}\label{eq:birks1}
\dLdx = \frac{S \; \dEdx}{1 + \kB\dEdx},
\end{equation}
where $\dEdx$ is the specific energy loss per unit path length, $\kB$ is the material-specific Birks' constant that describes the non-linear behavior of $L$ with $E$, and $S$ is the scintillation efficiency, i.e. the rate of photon production in an unquenched system. 

Electrons additionaly create Cherenkov light as they propagate, which is included in our analysis in order to accurately measure the total number of emitted scintillation photons, whereas $\alpha$ particles at nuclear energies are below the Cherenkov threshold. In Ref. \cite{vonKrosigk:2015aaa} it is shown that Birks' law provides a good fit to the behavior of the light output for $\alpha$ particles as a function of energy for a sample of LAB+PPO.

\subsection{Experimental Setup}\label{sec:light_yield_setup}

    The experimental setup for the light yield measurement uses a radioactive source, purchased from Spectrum Techniques \cite{spectrum_techniques}, placed in a cylindrical, UV-transparent acrylic vessel (AV) filled with the target liquid. A \isotope{Sr}{90} source is used for the $\beta$ emitter and \isotope{Po}{210} is used for the $\alpha$ emitter. The \isotope{Sr}{90} $\beta$ decays with Q-value \SI{546}{~\keV} to \isotope{Y}{90}, which then $\beta$ decays with Q-value \SI{2.28}{~\MeV} and half-life 64~hours \cite{sr90}. The \isotope{Po}{210} primarily decays via $\alpha$ emission with an energy of \SI{5.3}{~\MeV} to the ground state of \isotope{Pb}{206} with a very small branch (0.001\%) to an excited state that gives an \SI{803}{~\keV} $\gamma$ and \SI{4.5}{~\MeV} $\alpha$ \cite{po210}.

    The AV has a \SI{30}{~\milli\meter} diameter and is \SI{30}{~\milli\meter} tall. An inner cylindrical volume with a diameter of \SI{20}{~\milli\meter} and height of \SI{25}{~\milli\meter} is hollowed out and is used to hold the target material. The source is placed directly above the target material and rests on a ledge that is \SI{3.2}{~\milli\meter} thick. There is no acrylic between the target material and the source; however, there is a small gap of air, about \SI{2}{~\milli\meter} in width, between the target and source.

    The AV is placed on top of and optically coupled to a 1-inch square H11934-200 Hamamatsu photomultiplier tube (referred to at the trigger PMT) using Eljen Technology EJ-550 optical grease. The center of an R7081-100 10-inch Hamamatsu PMT (referred to as the light yield PMT) is placed \SI{10}{~\centi\meter} from the source. The signal from the trigger PMT is used to initiate the data aquisition (DAQ) system. The total charge collected at the light yield PMT is used to measure the light output of the target material by comparing the response to a detailed simulation model described in Sect.~\ref{sec:light_yield_sims}.

    The signals are digitized on a CAEN V1742 digitizer over a \SI{1}{~\volt} dynamic range, sampling at \SI{2.5}{~\giga\hertz} for 1024 samples, yielding waveforms that are \SI{409.6}{~\nano\second}{} long. The data is read out over USB using custom DAQ software \cite{Caravaca:2016ryf}, which produces \textsc{HDF5} files containing the waveforms.

\subsection{Waveform Processing}\label{sec:light_yield_waveform_analysis}

    The digitized waveforms from the trigger and light yield PMTs are processed by first calculating a per-waveform baseline using a \SI{120}{~\nano\second} window at the beginning of the trace. For the light yield PMT, events with PMT pulses are identified by selecting waveforms with peak values less than \SI{-5}{~\milli\volt}. The time corresponding to the peak of the PMT pulse is identified and the charge is calculated by integrating a dynamic window that extends \SI{10}{~\nano\second} before and \SI{100}{~\nano\second} after the sample associated with the peak. This window is selected to be long enough to include effectively all of the scintillation light produced by the \wbls{} \cite{Onken:2020pnv}.

\subsection{Simulating the Light Yield Setup}\label{sec:light_yield_sims}

\begin{figure}[b!]
    \centering
    \includegraphics[width=1.0\columnwidth]{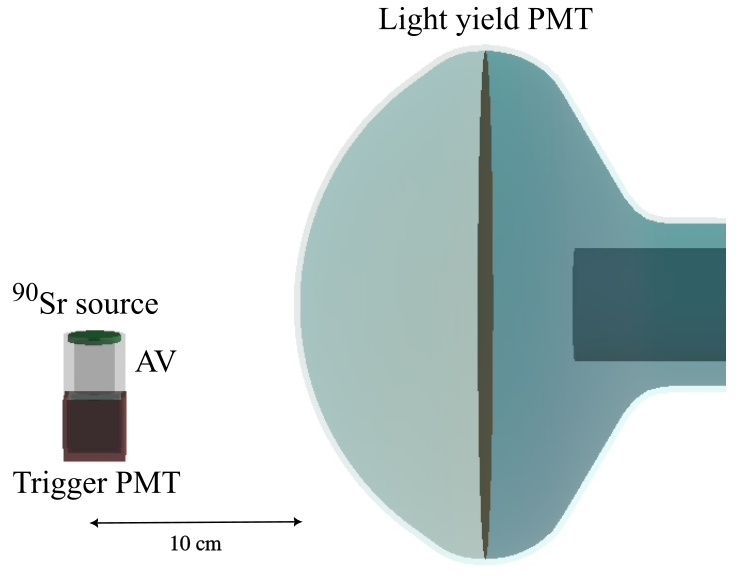}
    \caption{The light yield experimental setup with the \isotope{Sr}{90} source, as simulated in \textsc{RAT-PAC}.}
    \label{fig:exp_setup}
\end{figure}

    The light yield geometry, as simulated in \ratpac{}, is shown in Fig.~\ref{fig:exp_setup} as simulated in \textsc{RAT-PAC}.

    The PMT pulses are modeled by summing individual SPE pulses for each detected photoelectron (PE). The SPE pulses are modeled using a lognormal distribution:
\begin{equation}\label{eq:lognormal}
    f\pp{t} = B + \frac{Q}{(t - \tz)\sqrt{2\pi}\sigma} e^{-\frac{1}{2}\pp{\log\pp{\frac{t - \tz}{m}}/\sigma}^{2}},
\end{equation}
where $B$ is the baseline, $Q$ is the charge contained in the pulse, $\tz$ is arrival time of the pulse, and $m$ and $\sigma$ are shape parameters. These parameters are measured as described in Sect.~\ref{sec:light_yield_calibrations}. The individual pulses from each PE are summed linearly to create a single pulse, which is sampled at \SI{2.5}{~\giga\hertz} for 1024 samples. The output of the MC is a collection of \textsc{HDF5} files with identical structure to the data.

\subsection{Analysis Strategy}\label{sec:light_yield_analysis}

    Parameter estimation, whether for the light yield measurement or calibration purposes, is achieved by matching simulation to observed data. The use of a proper simulation ensures that optical effects and related physics, for example the production of Cherenkov light by electrons, are naturally accounted for. For any given parameter of the PMT or scintillation model, the parameter is tuned by producing simulations over small steps in that parameter, determining the optimal parameter estimate by minimizing the $\chi^{2}$ between the binned PMT charge observed in MC and data. For each parameter step, an ensemble of ten independent MC samples is produced, from which a mean $\chi^{2}$ and uncertainty may be computed. These $\chi^{2}$ estimates, as a function of the parameter value, are fit with a quadratic function in the neighborhood of the observed minimum, in order to identify the true minimum. The statistical uncertainty on the model parameter is determined by the $\Delta \chi^{2} = 1$ interval of the best-fit quadratic.

\subsection{Calibrations}\label{sec:light_yield_calibrations}

\begin{figure}[t!]
    \centering
    \includegraphics[width=1.0\columnwidth]{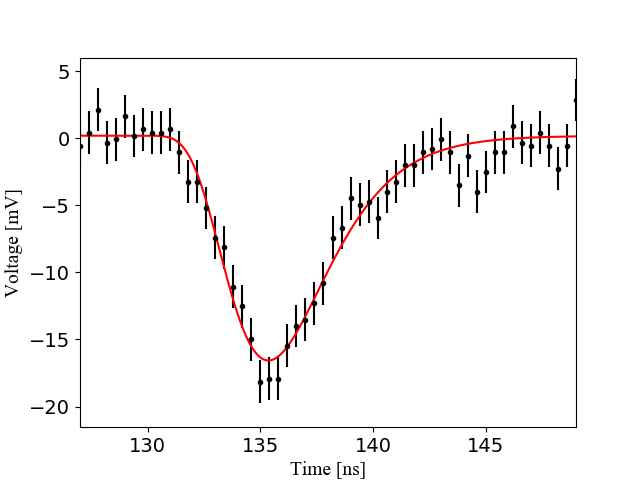}
    \caption{An example SPE PMT pulse with a lognormal fit (Eq.~\ref{eq:lognormal}).}
    \label{fig:waveform_with_lognormal_fit}
\end{figure}

    The SPE pulse shape and size of each PMT, and the photon detection efficiency of the light yield PMT must be calibrated in order to accurately model the data using the MC. The SPE charge distribution, in particular, is a critical model input, as the integrated charge is ultimately used to compare the MC with data. The pulse shapes and sizes are calibrated using a pulsed LED pointed at each PMT, respectively. The pulse to the LED is configured such that, in each pulse, the PMT detects primarily single photons. To extract the SPE pulse parameters, the recorded waveforms are fit with a lognormal function, given in Eq.~\ref{eq:lognormal}, performed by $\chi^{2}$-minimization. An example SPE pulse fit is shown in Fig.~\ref{fig:waveform_with_lognormal_fit}. The values of $m$ and $\sigma$ are binned, and fit with a Gaussian fit to determine a mean and standard deviation, which are input into the simulation model which accounts for pulse-to-pulse variation. The observed distribution of SPE charge, $Q$ is Gaussian for the trigger PMT, but not for the light yield PMT. To properly account for the non-Gaussian nature of the latter's charge distribution, a histogram of SPE charge values is used directly as input to the simulation and is sampled from when generating pulses in the light yield PMT.

\begin{figure}[t!]
    \centering
    \includegraphics[width=1.0\columnwidth]{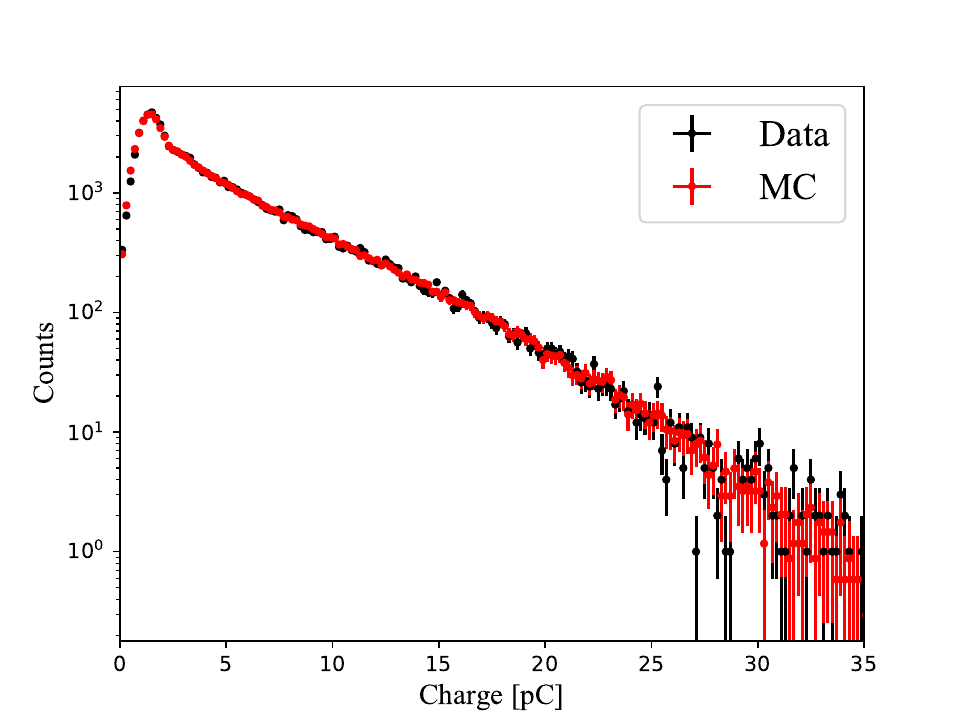}
    \caption{The charge distribution compared between data and MC for a water target and with an efficiency scaling in the MC of 63.5\%. The $\chi^{2}/\ndf$ is 162.3/175.}
    \label{fig:water_charge}
\end{figure}

    Having calibrated the SPE responses, the photon detection efficiency of the light yield PMT is calibrated using a pure Cherenkov source. The experimental setup described in Sect.~\ref{sec:light_yield_setup} is used, with water as the target material and a \isotope{Sr}{90} source providing $\beta$ radiation. This provides a well-understood source of light that can be used to tune the detection efficiency. This is implemented using global scale factor applied to the quantum efficiency curve, referred to as the ``efficiency scaling.'' This efficiency scaling accounts for unmodeled inefficiencies, such as the PE collection efficiency. The parameter step size used to sample the $\chi^{2}$ curve is 0.5\%. An example comparison of the data and MC using an efficiency scaling of 63.5\% is shown in Fig.~\ref{fig:water_charge}. Using the method outlined in Sect.~\ref{sec:light_yield_analysis}, a best fit value of $\pp{63.4 \pm 0.8 \; (\text{stat.})}\%$ is identified.

    Several sources of systematic uncertainty were investigated. Namely, we consider uncertainties arising from the integration window size, the light yield SPE charge model, and the geometry of the button source. Varying the low-level analysis cuts was determined to have a negligible impact. Systematics due to the wavelength spectra of the detected photons, which differ between Cherenkov and scintillation light, are not considered. This is expected to be a small systematic, as the wavelength spectra of detected Cherenkov and scintillation photons are similar. 

    The integration window length is varied between \SI{60}{~\nano\second} and \SI{140}{~\nano\second}{} and the maximum observed difference in the determined efficiency scaling of 2.4\% is conservatively taken as a two-sided uncertainty. The largest systematic uncertainty is from the SPE charge model for the light yield PMT, which we modify in the simulation from histogram sampling to a Gaussian response, with mean and width extracted from the LED data. This change leads to a 7.1\% difference in the best-fit efficiency scaling, which is also taken as a two-sided uncertainty. Lastly, within the plastic button source, the location of the \isotope{Sr}{90} is uncertain, and is specified with the corresponding uncertainties in the manufacturer drawings ($\pm$ 4\% on the radial distance to the edge and $\pm$ 33\% on the thickness of the plastic window between the \isotope{Sr}{90} and the bottom) \cite{spectrum_private}. The location of the source is varied in the MC within those constraints, leading to an asymmetric uncertainty of of $\pm^{4.8\%}_{4.1\%}$, driven primarily by the uncertainty in the distance between the active source and target material.

    The total systematic uncertainty is determined to be $\pm^{8.5\%}_{8.1\%}$. Including the statistical uncertainty, the best-fit efficiency scaling is $\pp{63.4 \pm ^{5.5}_{5.2}}\%$. The central value is used in the simulations of the \wbls{} measurements, and the uncertainty is propagated to the resulting light yield parameters.

\subsection{Measurement Methods}

\subsubsection{$\beta$ Measurements}\label{sec:beta_light_yield}

    Measurements using the \isotope{Sr}{90} source are primarily sensitive to the scintillation efficiency, $S$, due to the fact that $k_{B} \frac{dE}{dx} << 1$ for electrons. In this work, we tune the value of $S$ while holding $\kB$ fixed to the measured value for LAB+PPO. This is consistent with the methodology of similar previous measurements, such as those described in \cite{Caravaca:2020lfs,SNO:2020fhu}.

    We select a pure sample of \isotope{Y}{90} decays in data using a cut on the charge collected by the light yield PMT, as a proxy for the energy deposited in the scintillator. This removes the low energy \isotope{Sr}{90} component, requiring simulation of only the higher energy \isotope{Y}{90} decay. The value for the charge cut is determined by looking at the charge of the endpoint of the \isotope{Sr}{90} decays of the calibrated simulation output. This endpoint depends on the candidate scintillation efficiency, and we thus employ cut values well above the \isotope{Sr}{90} endpoint predicted even for unreasonably large values of $S$. The charge cut values used are \SI{10}{~\pico\coulomb}, \SI{30}{~\pico\coulomb} and \SI{50}{~\pico\coulomb} for 1\% \wbls, the 5\% \wbls, and the 10\% \wbls{} respectively.

    The simulations are run varying the value of $S$ in steps of \SI{5}{~photons/\MeV}, holding the value of $\kB$ fixed to \SI{0.074}{~\milli\meter/\MeV}, as measured for LAB+PPO in Ref. \cite{WanChanTseung:2011yv}. The value of $\kB$ for protons has been shown to be consistent between \wbls{} and oxygenated LAB+PPO \cite{Callaghan:2022ahi}. Nonetheless, varying this value at the level 10\% was found to have neglibible impact.

    The sources of systematic uncertainty considered are the calibrated scaling efficiency, the charge integration window length, and the value of the charge cut used to reject \isotope{Sr}{90} decays. The largest contribution is from the efficiency scaling, which in the simulation model is degenerate with $S$, with an uncertainty of $\pm^{8.5\%}_{8.1\%}$. The uncertainty from the integration window size must be reevaluated given the broader photon arrival time distribution associated with scintillation light than the Cherenkov light observed in water measurements. By varying the integration window length between 60 and \SI{140}{~\nano\second}{}, a conservative two-sided uncertainty of 4.5\% is observed. The value of the charge cut is varied around the endpoint of the \isotope{Sr}{90} (in ten steps of \SI{1}{~\pico\coulomb} above the nominal cut value) and the maximum change in the result is 0.8\%. The total systematic uncertainty is thus $\pm^{9.7\%}_{9.3\%}$.

\subsubsection{$\alpha$ Measurements}\label{sec:alpha_light_yield}

    Due to the larger $\alpha$ stopping power, measurements of the \isotope{Po}{210} source require consideration of $\kB$, as per Eq.~\ref{eq:birks1}. With only a single monoenergetic source, as used here, the two parameters of Birks' law are degenerate. Thus we extract the average number of scintillation photons produced, $\langle L \rangle$, from the best-fit MC model evaluation, and interpret this quantity in the context of Birks' model. To generate the results presented in this work, the value of $\kB$ was held constant to the value measured for LAB+PPO \cite{vonKrosigk:2015aaa}, and the value of $S$ was tuned as described above. It was verified that equivalent results for $\langle L \rangle$ are achieved if, instead, $S$ is held constant and the value of $\kB$ is varied. Because over 99.9\% of \isotope{Po}{210} decays produce a single monoenergetic $\alpha$, we do not apply any cut on the charge to select the signal. The $\alpha$ particles from the \isotope{Po}{210} decays deposit an average of \SI{4.8}{~\MeV} in the \wbls{}, as they lose some energy in the air between the source and the target.

    The explicit sources of systematic uncertainty considered are the calibrated scaling efficiency, the charge integration window length, and an uncertainty arising from the sample geometry, namely the height of the small air-gap between the source housing and sample surface. The largest contribution is again the $\pm^{8.5\%}_{8.1\%}$ uncertainty on the value of the efficiency scaling. The same conservative two-sided systematic of 4.5\% is determined by varying the integration window length. The height of the small gap of air between the source and the sample impacts the total amount of energy deposited by $\alpha$ particles in the target material much more than electrons, due to the difference in stopping powers. The nominal height of the air gap is measured with calipers, and the weight of the sample is standardized to ensure the same amount of material is used in each measurement. With these procedures we expect the height of the gap of air to be consistent across all measurements to within \SI{0.3}{~\milli\meter}{}. To be conservative, the height of the gap of air is varied in simulation by $\pm \SI{1}{~\milli\meter}$, and the observed change in the best fit light yield is $\pm^{2.2\%}_{3.1\%}$.

    We compute an additional uncertainty associated with indeterminate deficiencies in the simulation model, apparent when comparing the data and MC, as presented in Sect.~\ref{sec:light_yield_results_alphas}. This is quantified by determining a scaling to the uncertainty on $\chi^{2}$ estimates sampled from MC, such that the minimum satisfies $\chi^{2}/\ndf = 1$. This scaling is then interpreted as a residual fractional uncertainty. Quantitatively, this increases the total systematic uncertainty on the measurement of the 10\% sample by approximately a factor of 3, and is negligible for the 1\% and 5\% samples. The total systematic uncertainties on the measurements of the 1\%, 5\% and 10\% samples are $\pm^{9.9\%}_{9.8\%}$, $\pm^{9.9\%}_{9.8\%}$, and $\pm^{29.7\%}_{29.4\%}$, respectively.

\subsection{Results}\label{sec:light_yield_results}

\subsubsection{$\beta$ Measurements}\label{sec:light_yield_results_betas}

\begin{figure*}[t!]
    \centering
    \includegraphics[width=0.65\columnwidth]{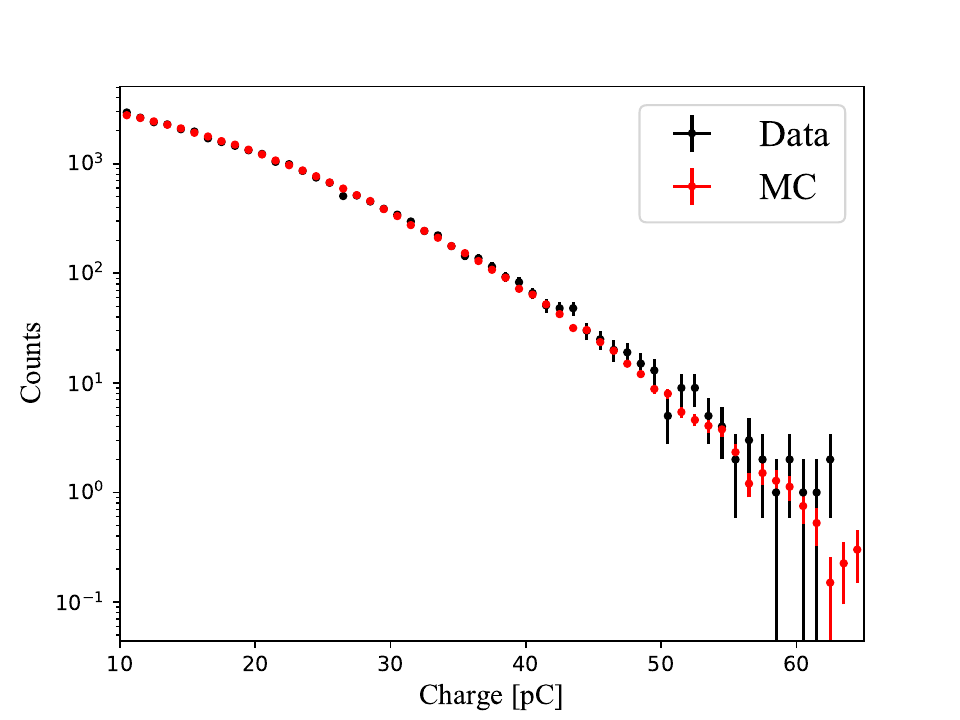}
    \includegraphics[width=0.65\columnwidth]{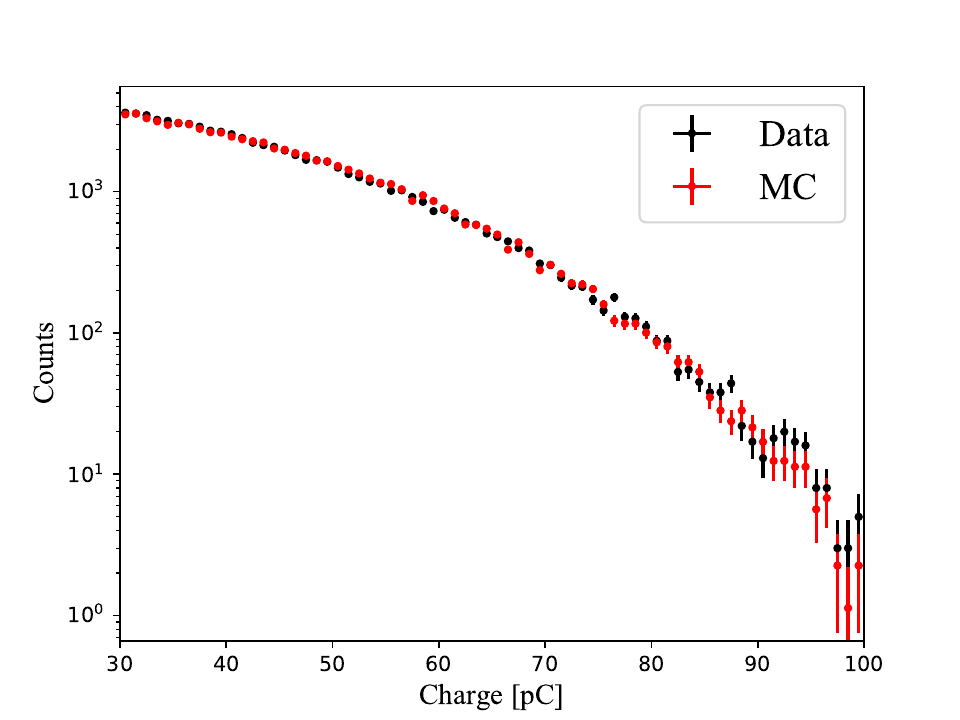}
    \includegraphics[width=0.65\columnwidth]{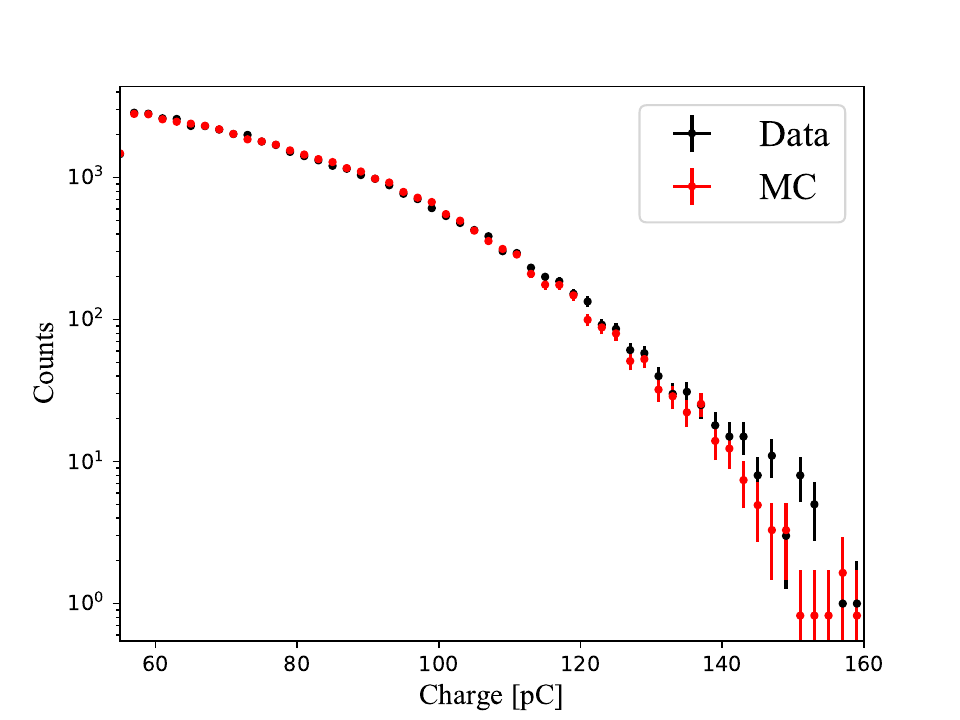}
    \caption{The light yield PMT charge distributions compared between the $\beta$ data and MC for the 1\% \wbls{} with an $S$ of \SI{257}{~photons/\MeV} (left), the 5\% \wbls{} with an $S$ of \SI{754}{~photons/\MeV} (center), and the 10\% \wbls{} with an $S$ of \SI{1380}{~photons/\MeV} (right). The values of $\chi^{2}/\ndf$ are 53.9/53, 110.4/73, and 51.4/52 respectively.}
    \label{fig:wbls_beta_charge}
\end{figure*}

    The light yield PMT charge distributions of simulations using the best fit values of $S$ are compared to the data in Fig.~\ref{fig:wbls_beta_charge}, exhibiting good agreement, with $\chi^{2}/\ndf$ $\sim$ 1. Table~\ref{tab:light_yield_results} lists the measured values of $S$, which are consistent with the values previously reported in Ref. \cite{Caravaca:2020lfs}.

\begin{table}[t!]
    \centering
    \caption{Measured scintillation efficincies using \isotope{Y}{90} $\beta$-decays (Q-value of \SI{2.28}{~\MeV}) in \wbls{}, assuming Birks' law with $\kB$ as measured for LAB+PPO \cite{WanChanTseung:2011yv}.}
    \begin{tabular}{c|c} \hline \hline
         Material & $S$~[photons/\MeV{}] \\ \hline
         1\% \wbls{}  & 257 $\pm 4$ (stat.) $\pm^{25}_{24}$ (syst.) \\ [0.75ex]
         5\% \wbls{}  & 754 $\pm 10$ (stat.)  $\pm^{73}_{70}$ (syts.) \\ [0.75ex]
         10\% \wbls{} & 1380 $\pm 14$ (stat.) $\pm^{134}_{128}$ (syst.) \\ [0.75ex] \hline
    \end{tabular}
    \label{tab:light_yield_results}
\end{table}

\subsubsection{$\alpha$ Measurements}\label{sec:light_yield_results_alphas}

\begin{figure*}[t!]
    \centering
    \includegraphics[width=0.65\columnwidth]{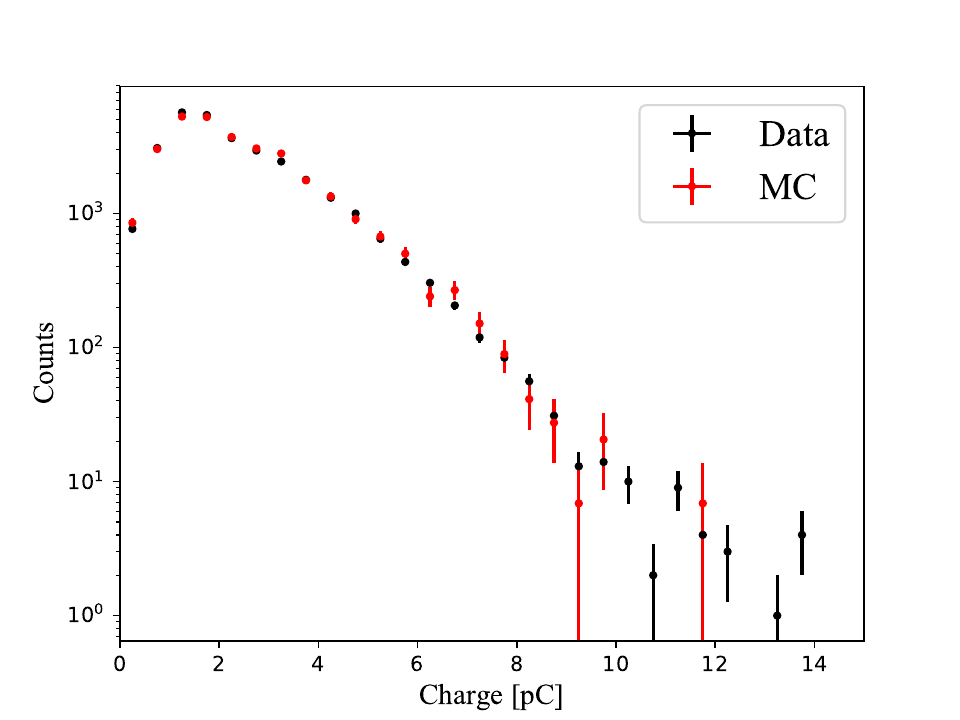}
    \includegraphics[width=0.65\columnwidth]{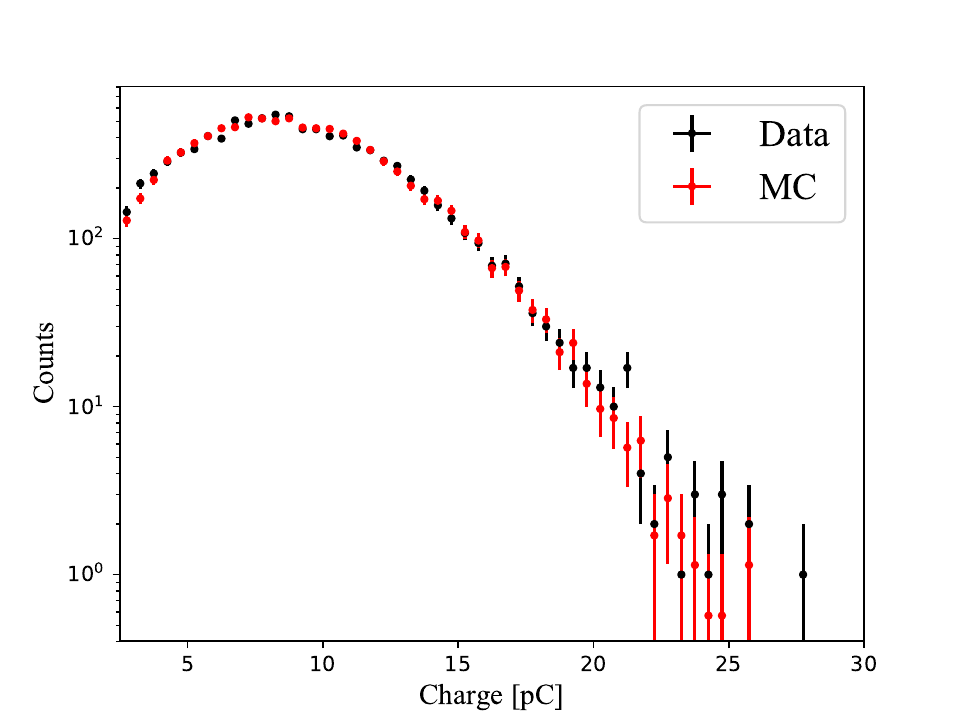}
    \includegraphics[width=0.65\columnwidth]{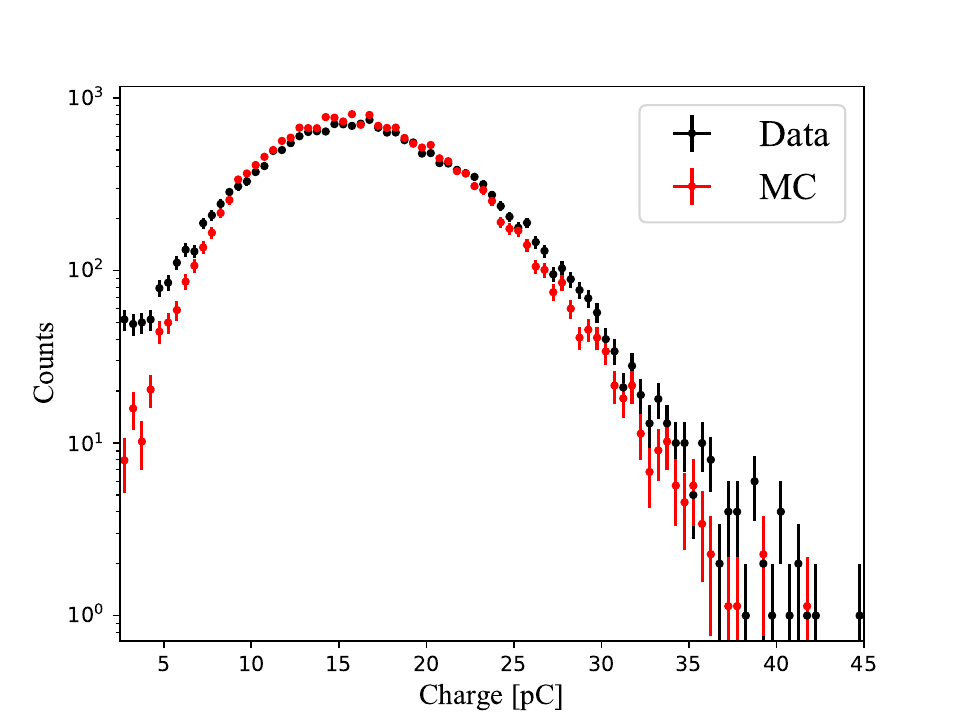}
    \caption{The light yield PMT charge distributions compared between the $\alpha$ data and MC for the 1\% \wbls{} with an S of \SI{125}{~photons/\MeV}, the 5\% \wbls{} target with an $S$ of \SI{640}{~photons/\MeV} (center), and the 10\% \wbls{} target with an $S$ of \SI{1220}{~photons/\MeV} (right). The values of $\chi^{2}/\ndf$ are 21.7/21, 38.1/46, and 289.7/72 respectively.}
    \label{fig:wbls_alpha_charge}
\end{figure*}

    The light yield PMT charge distributions of simulations using the best-fit model parameters are compared to the data in Fig.~\ref{fig:wbls_alpha_charge}, exhibiting generally good agreement, with $\chi^{2}/\ndf$ $\sim$ 1 (with the exception of the 10\% WbLS, discussed later). The uncertainties shown on the data points for both data and MC include only statistical uncertainty, and the $\chi^{2}$ values are calculated over the ranges shown. The 10\% \wbls{} data is broader than predicted by the simulation model, which contributes to the systematic uncertainty on the measurement as described in Sect.~\ref{sec:alpha_light_yield}.

\begin{figure}[ht!]
    \includegraphics[width=1.0\columnwidth]{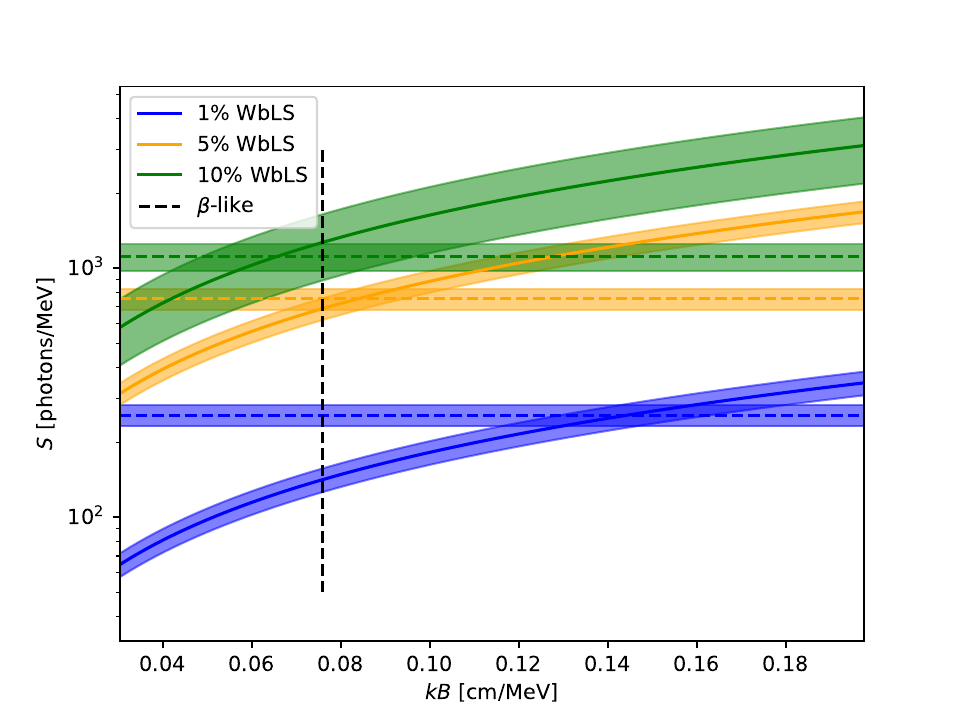}
    \caption{Summary of constraints on Birks' law parameters in \wbls{}. Solid lines show parameter constraints for $\alpha$ particles. Colored dashed lines show the measured scintillation efficiencies for electrons, and the black dashed line denotes the value of Birks' constant measured for LAB+PPO, which in this work is assumed for \wbls{}. Uncertainty bands contain both statistical and systematic uncertainties.}
    \label{fig:birks_alphas}
\end{figure}

    The average numbers of scintillation photons produced in the best-fit simulations are listed in Table~\ref{tab:light_yield_results_alphas}, and the corresponding constraints on the parameters of Birks' law are shown in Fig.~\ref{fig:birks_alphas}, with the assumed Birks' constant and measured scintillation efficiencies for electrons overlaid. It is observed that there are consistent model parameters for higher scintillator loadings, namely the 5\% and 10\% samples, whereas the 1\% sample produces less light than predicted by electron-like Birks quenching. A possible explanation is the breakdown of simple ionization quenching in low scintillator loadings, where relatively little energy is deposited directly to scintillating micelles. Such a departure from Birks' law would manifest differently for electrons and $\alpha$s, owing to their different stopping ranges.

\begin{table}[t!]
    \centering
    \caption{Average numbers of scintillation photons produced by \SI{4.8}{~\MeV} $\alpha$ particles in \wbls{}.}
    \begin{tabular}{c|c} \hline \hline
         Material & $\langle L \rangle$~[photons] \\ \hline
         1\% \wbls{}  & 58 $\pm$ 2 (stat.) $\pm$ 6 (syst.)  \\
         5\% \wbls{}  & 281 $\pm$ 3 (stat.) $\pm$ 28 (syst.) \\
         10\% \wbls{} & 516 $\pm$ 8 (stat.) $\pm$ 153 (syst.) \\ \hline
    \end{tabular}
    \label{tab:light_yield_results_alphas}
\end{table}

%% file: timing.tex
\section{Emission Timing}\label{sec:timing}
    The scintillation time profile of two-component systems, containing a primary solvent and secondary fluor, are typically modeled as a sum of exponentially-decaying terms, each modified by a common ``rise time,'' which accounts for excitation of fluor molecules via non-radiative energy transfer from solvent molecules, as described in \cite{birks}:
    \begin{align}
        S\pp{t} = &\sum_{i=1}^{N}
                    A_{i}
                    \frac{\eto{-t/\tau_{i}} - \eto{-t/\tR}}
                    {\tau_{i} - \tR}, \\
        &\sum_{i=1}^{N} A_{i} = 1,
    \label{eq:emission_time_profile}
    \end{align}
    where $t$ is the time of photon emission relative to solvent excitation, $\tau_{i}$ are the lifetimes of the $N$ observable decay modes, $\tR$ is the rise time, and $S$ is normalized so that $\int_{0}^{\infty} \differential{t'}\;S\pp{t'} = 1$.

    The theory of organic scintillators usually associates the observed scintillation with primary excitation and ionization of $\pi$-electrons \cite{birks}. Direct excitation is into a singlet state; in the case of full ionization, electron pairs recombine preferentially into a relatively long-lived triplet state \cite{birks}. The larger stopping power for $\alpha$s, relative to electrons, thus translates to a higher proportion of triplet states, and hence a generically slower scintillation time profile. This is the basis for timing-based PID. We describe below measurements of the scintillation time profiles of $\alpha$s for three materials: 5\% \wbls{}, 10\% \wbls{}, and LAB+PPO. The responses of these materials to electrons was reported in \cite{Kaptanoglu:2021prv}. The time profile of 1\% \wbls{} was not measured in this work, as robust data-taking was impractical due to its low light yield under $\alpha$ radiation, coupled with the unsealed vessel design and low coincidence rate intrinsic to the method described below.

\subsection{Experimental Setup}

\begin{figure}[t!]
    \centering
    \includegraphics[width=1.0\columnwidth]{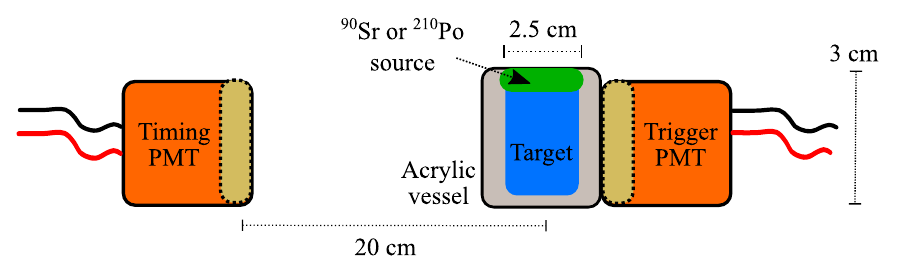}
    \caption{Diagram of the experimental setup used to measure emission timing. The trigger PMT is coupled directly to an acrylic vessel containing the sample under study, and operates in the multiphoton regime. A second PMT is located at a \SI{20}{\si{\centi}\meter} standoff, and operates in the single-photon regime in low coincidence with the trigger PMT. There is no acrylic interface between the radiation source housing and the sample volume.}
    \label{fig:timing_setup}
\end{figure}

    The target material, acrylic vessel, and radioactive source were arranged as described in Sect.~\ref{sec:light_yield_setup}. The acrylic vessel was viewed by two Hamamatsu H11394-200 PMTs, the signals from which were digitized using a CAEN V1742 operating at \SI{5}{\giga\hertz{}}. Digitization was triggered on the rising edge of one PMT coupled to the acrylic vessel; the other was placed at a distance of $\sim \SI{20}{\si{\centi}\meter}$, operating in the single-photon regime. The experimental setup is shown in Figure \ref{fig:timing_setup}.

    The LAB+PPO was sparged with nitrogen gas to remove dissolved oxygen in the sample, which is known to affect the scintillation time-profile \cite{OKeeffe:2011dex}. This is not performed for the \wbls{} samples, whose time-profiles are not impacted by the removal of oxygen \cite{Kaptanoglu:2021prv}.

\subsection{Waveform Processing}\label{sec:timing_waveform_analysis}

    The digitizer produces waveforms of a total length of \SI{204.8}{\nano\second}. For each waveform, the baseline is calculated using a $\SI{15}{\nano\second}$ window preceding the prompt light arrival and the total charge collected is computed by integrating a $\SI{120}{\nano\second}$ window following the arrival of the prompt light. The time of the photon for the SPE timing PMT is determined by applying a constant-fraction discriminator (CFD) to the waveform. A fractional threshold of 60\% is used with linear interpolation between the two samples where the pulse crossed threshold. The output of applying the CFD produces $t_{PMT}$. For the trigger PMT, a time value, $t_{trig}$ is assigned at the $\SI{3}{\milli\volt}$-threshold-crossing time. This threshold corresponds to roughly ${1}/{3}$ of a PE and is used to determine the time of the first photon detected in the trigger PMT, as discussed in Sect.~\ref{sec:timing_analysis}. The quantity $\Delta t = t_{PMT} - t_{trig}$ is used to extract the scintillation emission time, as discussed in the following section.

\subsection{Analysis Strategy}\label{sec:timing_analysis}

    The emission time profile of each sample under study is determined by fitting an analytic model to the observed time profile, with particular attention paid to the form of the system response, which is determined by the trigger scheme. As detailed below, this entails both modeling the effect on the system response of the occupancy levels of the trigger PMT, and the independent determination of the occupancy levels. Taken as a fixed input for the time profile analysis, this allows for the dependence on the system response of the scintillation emission time profile to be modeled during fitting, which improves the modeling of the scintillation rise time.

    The distribution of observed, or ``measured,'' SPE photodetector times, relative to the system trigger, can conceptually be written as
    \begin{equation}
        M\pp{t}
            = S\pp{t} \otimes P\pp{t} \odot T\pp{t} \otimes \delta\pp{t - \tz}
    \label{eq:observed_timing}
    \end{equation}
    where $S$ is the scintillation emission profile; $P$ is the response function of the photodetector; $T$ is the {\it trigger profile}, i.e. the distribution of times between data acquisition triggering and the deposition of energy into the sample; $\tz$ is an overall system delay; $\otimes$ denotes convolution, i.e. $F \otimes G = \int_{-\infty}^{+\infty} \differential{t'} F\pp{t'} G\pp{t - t'}$; $\odot$ denotes truncated anticonvolution, i.e. $F \odot G = \int_{0}^{+\infty} \differential{t'} F\pp{t + t'} G\pp{t'}$; and all operators are applied in the order reading from left to right. The anticonvolution with the trigger profile must be truncated to respect causality, as the system cannot trigger before any photons are detected. The PMT response function $P$ is approximately Gaussian \cite{h11934_datasheet}, but the trigger profile $T$ is, in general, asymmetric: with the single-photon trigger times defined in Sect.~\ref{sec:timing_waveform_analysis}, $T$ can be written in terms of the first order-statistic (see, e.g., \cite{OrderStatistics} for mathematical context) of the scintillation emission profile $S$ -- that is, given $n$ photons detected by the trigger PMT, the probabilty density function of the time the first detected photon. This can be written analytically: if $Q\pp{t}$ is the cumulative distribution function associated with the probability distribution function $S\pp{t}$, then the first order-statistic of $S$ given $n$ detected photons is
    \begin{equation}
        S_{1}^{\pp{n}}\pp{t} = n S\pp{t} \pp{1 - Q\pp{t}}^{n-1}.
    \end{equation}
    In practice, the number of photons collected by the trigger PMT is not constant, but varies event-to-event. If $W_{i}$ is the fraction of events in which the trigger PMT detected $i$ photons, then the trigger profile can be written as
    \begin{align}
        T\pp{t} = P'\pp{t} \otimes
        &\sum\limits_{n=1}^{\infty} W_{n} S_{1}^{\pp{n}}\pp{t}, \\
        &\sum\limits_{n=1}^{\infty} W_{n} = 1,
    \end{align}
    where $P'\pp{t}$ is the response function of the trigger PMT. The occupancy fractions (or occupancy spectrum) $\{W_{i}\}$ are determined by calibrating the SPE charge of the trigger PMT, as described in Sect.~\ref{sec:timing_calibration}, and fitting the observed multi-PE charge spectrum during data-taking, as described in Sect.~\ref{sec:occupancy_spectrum}. With this formulation, the anticonvolution in Eq.~\ref{eq:observed_timing} is then evaluated numerically. In this work, we allow for two deexcitation modes (i.e. $N = 2$ in Eq.~\ref{eq:emission_time_profile}), and model the combined photodetector response of the two PMTs, $P \otimes P'$, as a Gaussian with $\sigma = \SI{163}{\pico\second}$, corresponding to the combined response of two identical devices operating at manufacturer specification \cite{h11934_datasheet}. The dependence of the trigger profile $T\pp{t} = T\pp{t; \tR, \tau_{1}, \tau_{2}, A_{1}}$ on the candidate emission time profile is included.

\subsection{Trigger PMT Calibration}\label{sec:timing_calibration}

    In order to determine the distribution of trigger PMT occupancies, as described in Sect.~\ref{sec:occupancy_spectrum}, this single-photon charge response of the trigger PMT must be parameterized and calibrated. This was accomplished using single-photon pulses provided by a blue LED. The LED was pulsed to establish a low coincidence rate (below 5\%) in order to suppress multiphoton contamination. The SPE charge was determined using an integration window of $\SI{18}{\nano\second}$, which captures the full width of SPE pulses.

\begin{figure}[t!]
    \centering
    \includegraphics[width=0.45\textwidth]{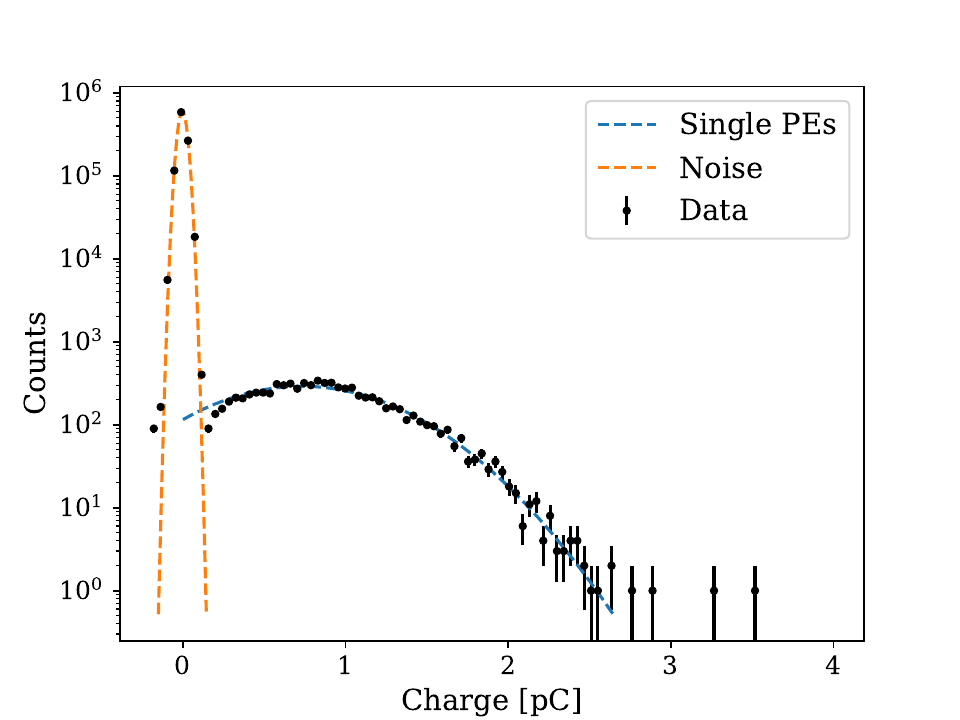}
    \caption{Distributions of trigger PMT charge when flashing an LED, calculating using $\SI{18}{\nano\second}$ integration window.}
    \label{fig:timing_calibration}
\end{figure}

    The distribution of SPE charge is shown in Fig.~\ref{fig:timing_calibration}, and is fit with a two-Gaussian model, corresponding to pure noise and a single-photon signal. This simple model was found to adequately describe the observed charge distribution with best-fit signal parameters of $\mu = \pp{0.731 \pm 0.009}\;\SI{}{\pico\coulomb}$ and standard deviation $\sigma = \pp{0.538 \pm 0.007}\;\SI{}{\pico\coulomb}$. These values are consistent with cross-check calibrations performed using longer integration lengths, though the latter provide weaker constraints due to the higher level of electronic noise included in the integration.

\subsection{Determining the Trigger Occupancy Spectrum}\label{sec:occupancy_spectrum}

    The trigger occupancy spectrum, i.e. the distribution of the number of photons detected by the trigger PMT, can be extracted from the trigger PMT charge spectrum collected during data-taking, given models for both the SPE charge response and charge summation. As discussed in Sect.~\ref{sec:timing_calibration}, the distribution of SPE charge $q$ is approximately a Gaussian function $G\pp{q; \mu,\sigma}$, with mean $\mu$ and standard deviation $\sigma$. We assume that SPE pulses add perfectly linearly, so that the charge response associated with an $n$-PE pulse is also a Gaussian $G\pp{q; n\mu,\sqrt{n}\sigma}$ in accordance with the central limit theorem. Given an observed distribution $P\pp{q}$ of charge $q$, we can write
    \begin{equation}
        P\pp{q; \{W_{i}\}} = \sum\limits_{n=1}^{\infty} W_{n} G\pp{q; n\mu,\sqrt{n}\sigma}
    \label{eq:mpe_charge}
    \end{equation}
as a sum over multi-PE charge distributions, with $W_{n}$ defined as in Sect.~\ref{sec:timing_analysis}.

    Regarding the fractions $\{W_{i}\}$ as free parameters, Eq.~\ref{eq:mpe_charge} can be fit to the distribution of trigger PMT charge observed during data-taking. The fit is formulated as a least-square-difference minimization between the binned charge data and model. In practice, the infinite sum must be truncated at some finite maximum mode, $n_{\text{max}}$. In developing the results presented in this work, several choices for $n_{\text{max}}$ were used, and it was found that the fit results are robust to changes to the exact value, as long as the mean charge of the maximum mode, $n_{\text{max}}\mu$, is well-above the endpoint of the observed charge distribution. The high dimensionality associated with the generality of this model (choosing $n_{\text{max}} = 200$, for example, requires a 200-dimensional minimization) is dealt with by evaluating the model using the \jax{} \cite{jax2018github} library, which allows the cost function, $C_{0}$, to be differentiated analytically via automatic differentiation. This allows for efficient minimization using gradient descent, without the presence of the numerical error which accrues when approximating the gradient using finite difference methods. In this work, we employ a basin-hopping algorithm, in which simulated annealing is applied to repeated local gradient-descent-based minimizations.

\begin{figure*}[t!]
    \centering
    \includegraphics[width=0.330\textwidth]{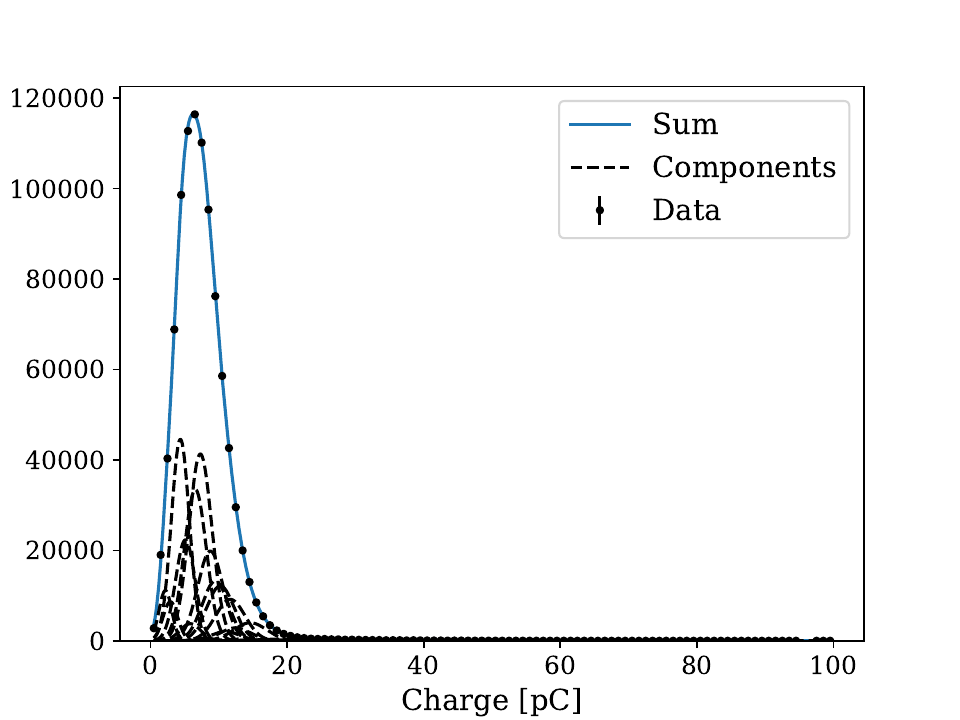}
    \includegraphics[width=0.330\textwidth]{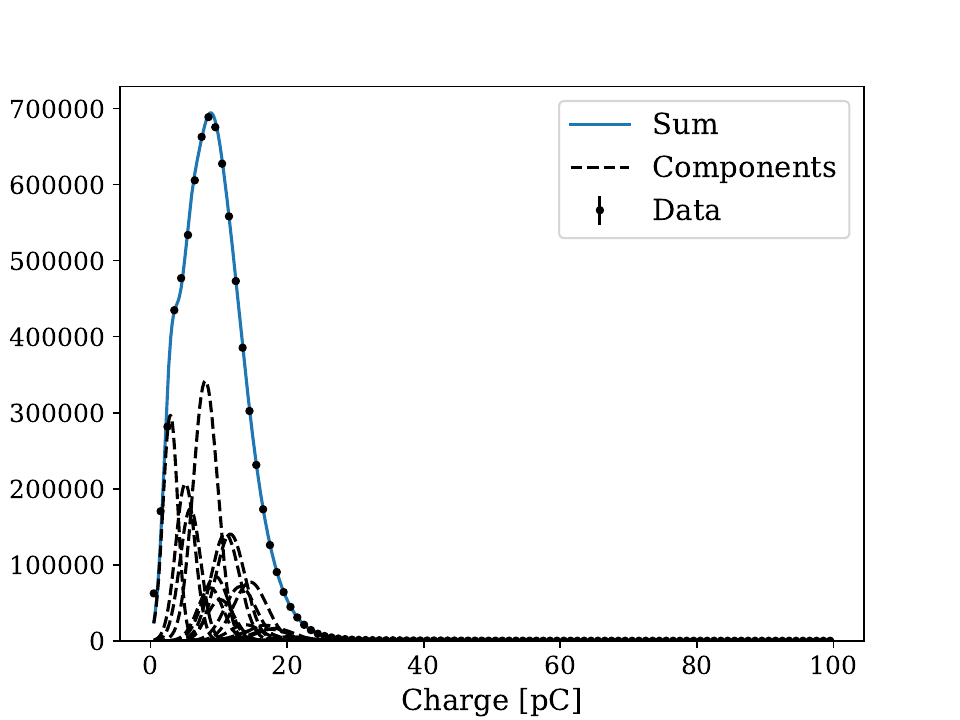}
    \includegraphics[width=0.330\textwidth]{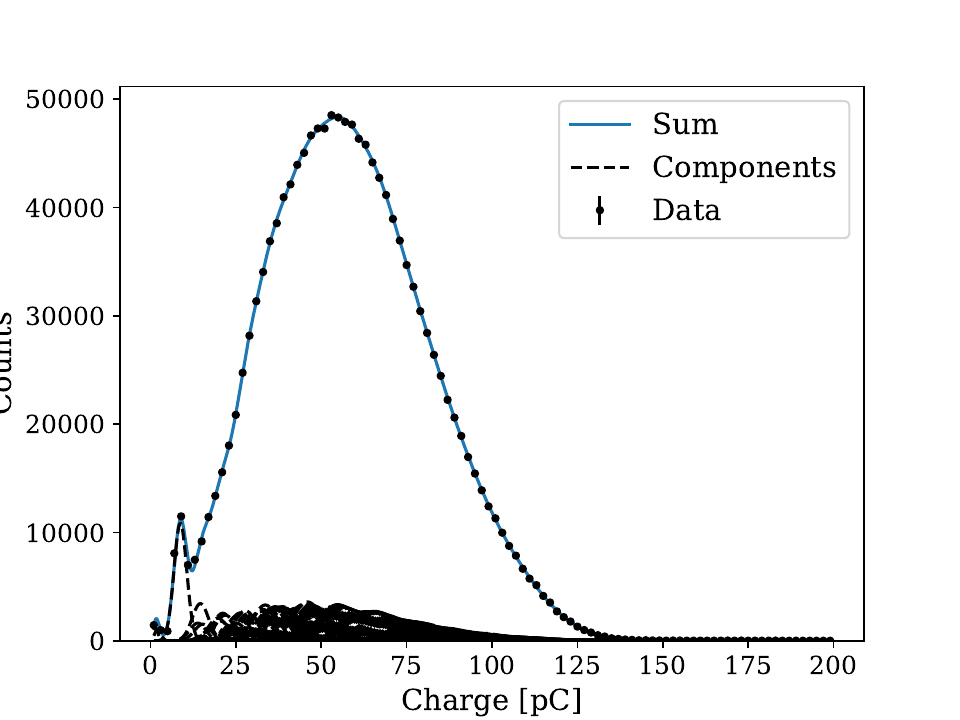}
    \caption{Distribution of trigger PMT charge during $\alpha$ data-taking, with best-fit multi-PE charge model overlaid, for 5\% \wbls{} (left), 10\% \wbls{} (middle), and LAB+PPO (right).}
    \label{fig:charge_spectra}
\end{figure*}

\begin{figure*}[t!]
    \centering
    \includegraphics[width=0.330\textwidth]{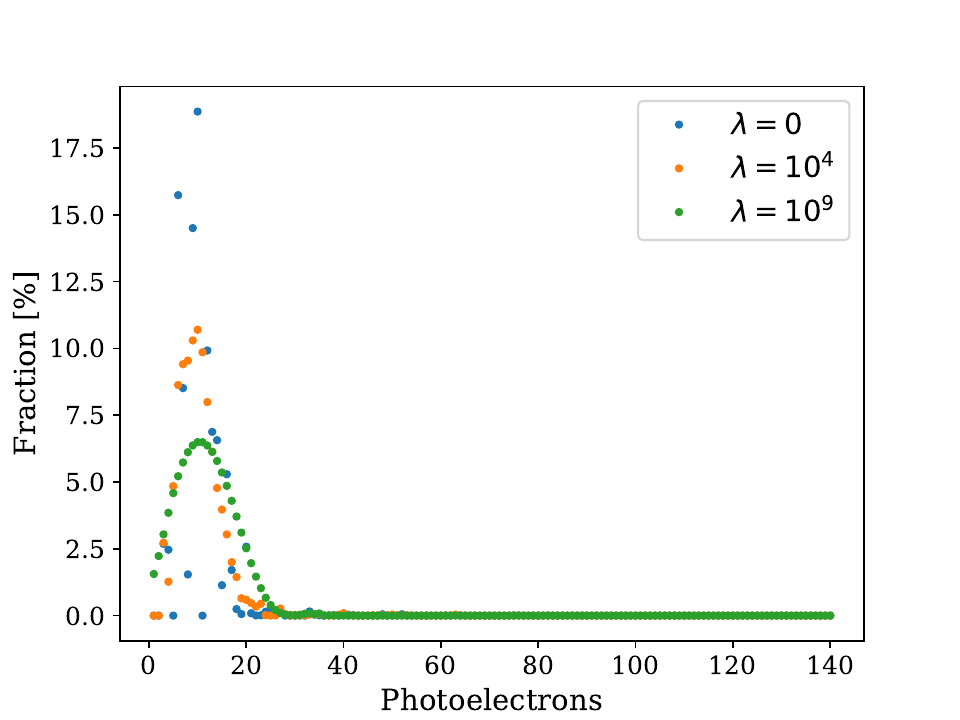}
    \includegraphics[width=0.330\textwidth]{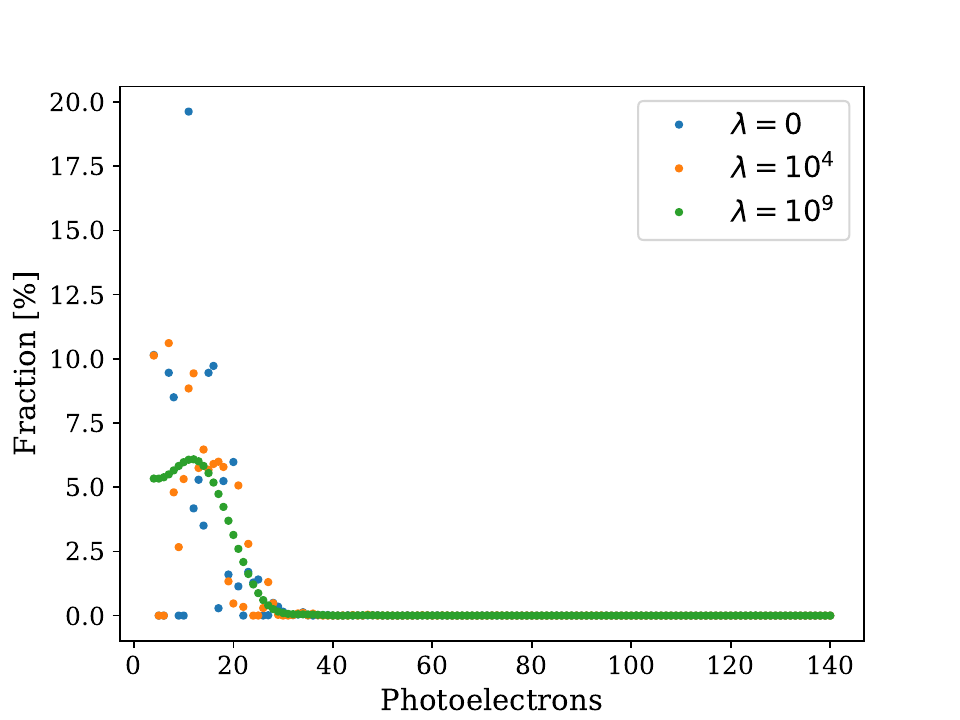}
    \includegraphics[width=0.330\textwidth]{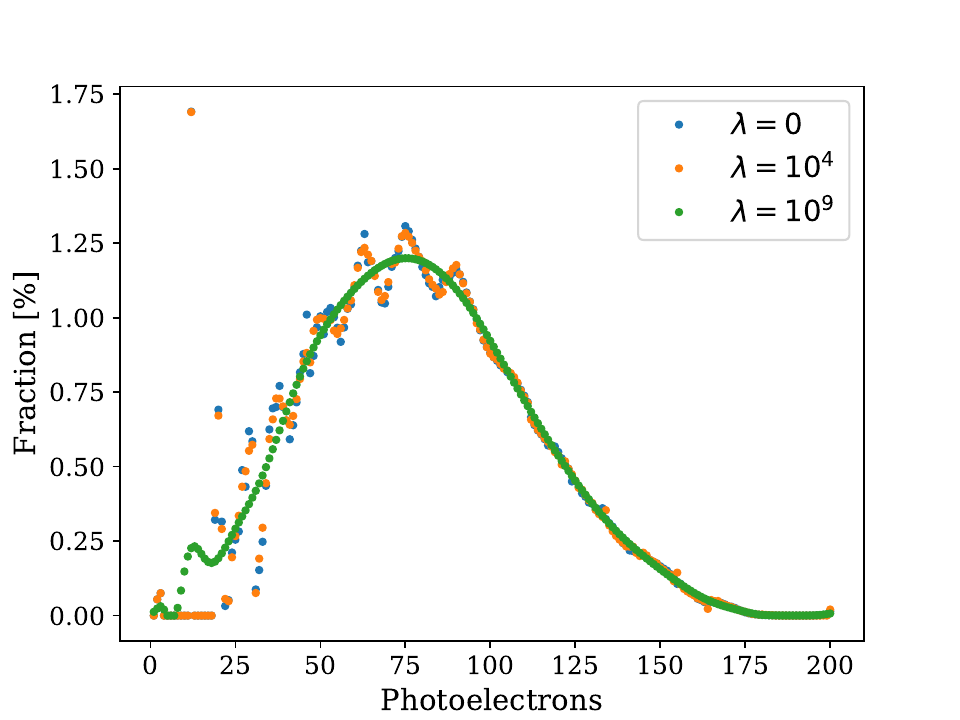}
    \caption{Best-fit spectra of trigger PMT occupancies during $\alpha$ data-taking, for 5\% \wbls{} (left), 10\% \wbls{} (middle), and LAB+PPO (right), under different modality-penalization strengths.}
    \label{fig:occupancy_spectra}
\end{figure*}

    It was observed that the best-fit occupancy spectra exhibit oscillatory behavior, wherein the preferred occupancy varies between several local maxima and minima, as opposed to a unimodal distribution, as might be expected from a monoenergetic source. This is believed to be an artifact of imperfect minimization owing to the combinatorics of high-dimensality and large correlations between neighboring occupancy modes, intrinsic to the general model defined by Eq.~\ref{eq:mpe_charge}: because neighoring occupancies are nearly-degenerate, there are many minima, which we term ``impostor minima,'' in which one occupancy representative of its neighborhood contains an excess weighting, but only one minima where the weighting is correctly allocated between neighboring modes. Whether the minima identified are sufficient to properly model the trigger response was studied by manually enforcing unimodality in the occupancy spectrum via a constraint term in the cost function, which penalizes spectra according to the difference between neighboring modes. That is, we instead minimize a penalized cost function
    \begin{equation}
    C_{\lambda}\pp{\{W_{i}\}}
        = C_{0} + \lambda \sum\limits_{n=1}^{\infty} \pp{W_{n+1} - W_{n}}^{2},
    \end{equation}
where the penalization strength $\lambda$ controls level of constraint on the spectrum shape: $\lambda = 0$ results in the ordinary least-squares cost function defined previously, whereas in the limit $\lambda \rightarrow \infty$ the preferred occupancy spectrum is uniform across all allowed modes. The scintillation time profile results presented in this work are robust to machine precision for different choices of $\lambda$ spanning 10 orders of magnitude, indicating that the small difference between the impostor and true global minima is below the sensitivity of our timing data to the trigger profile.

    Observed trigger charge spectra and best-fit models are shown in Fig.~\ref{fig:charge_spectra}. Best-fit occupancy spectra, for a selection of penalization strengths, are shown in Fig.~\ref{fig:occupancy_spectra}.

\subsection{Results}\label{sec:timing_results}

\begin{figure*}[t!]
    \centering
    \includegraphics[width=0.330\textwidth]{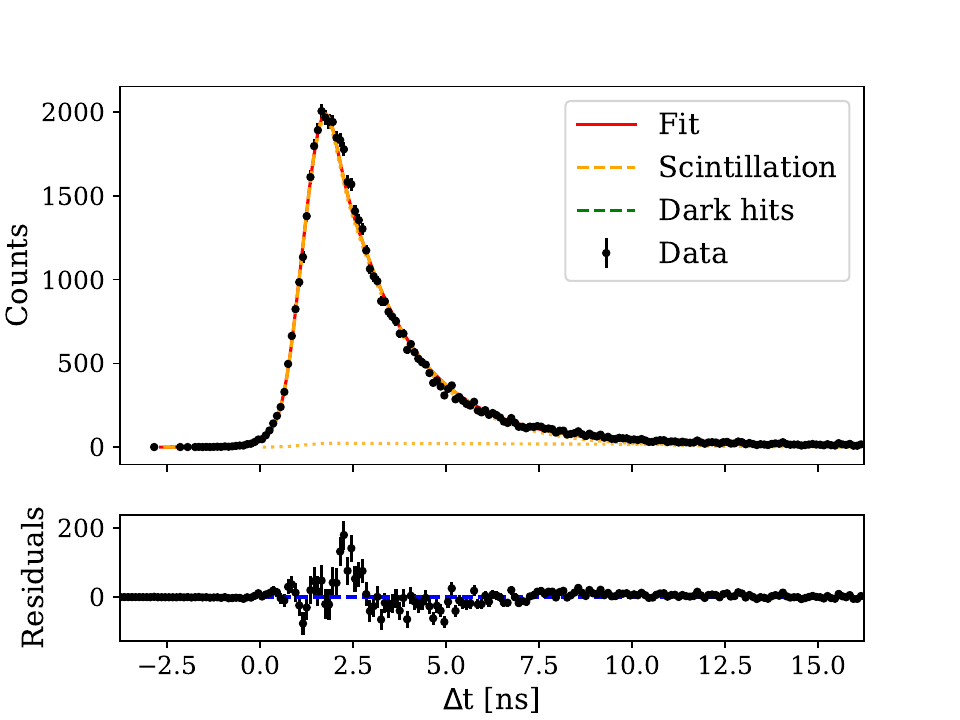}
    \includegraphics[width=0.330\textwidth]{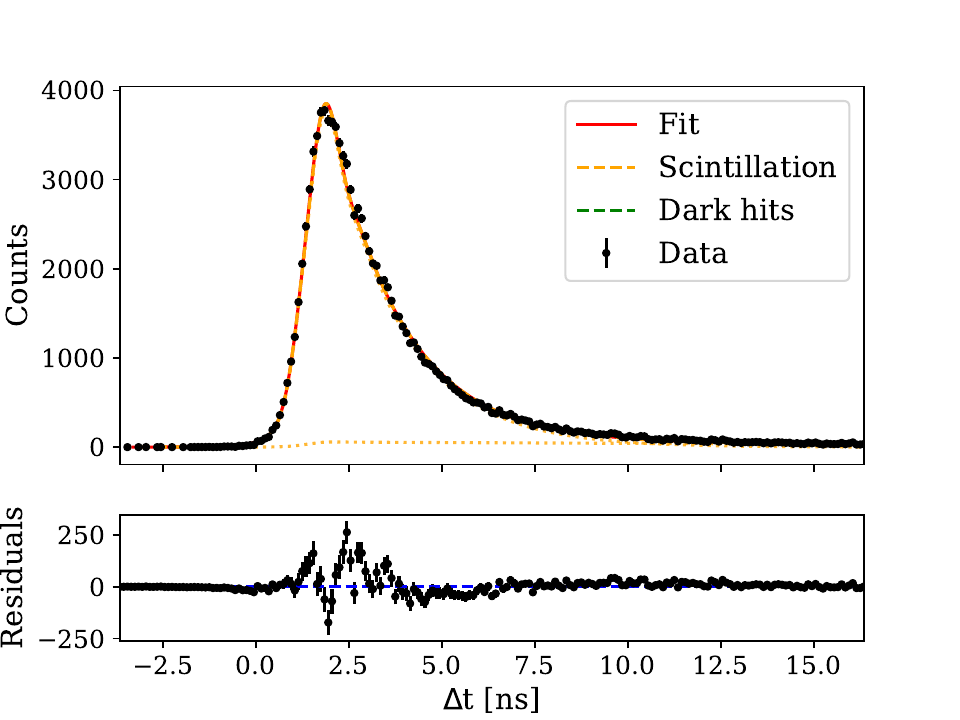}
    \includegraphics[width=0.330\textwidth]{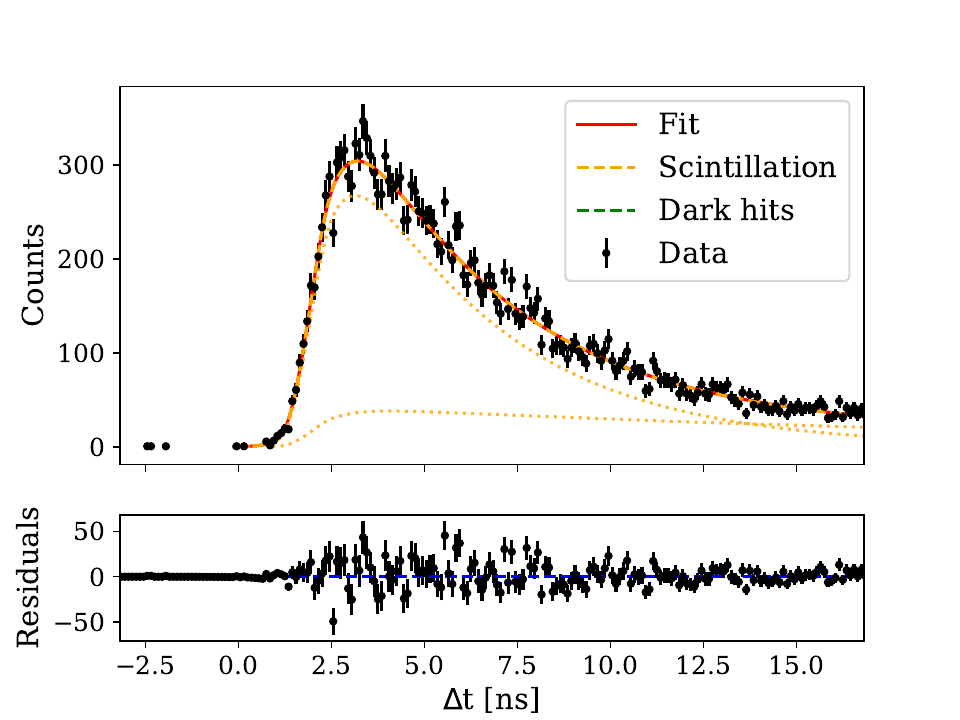}
    \\
    \includegraphics[width=0.330\textwidth]{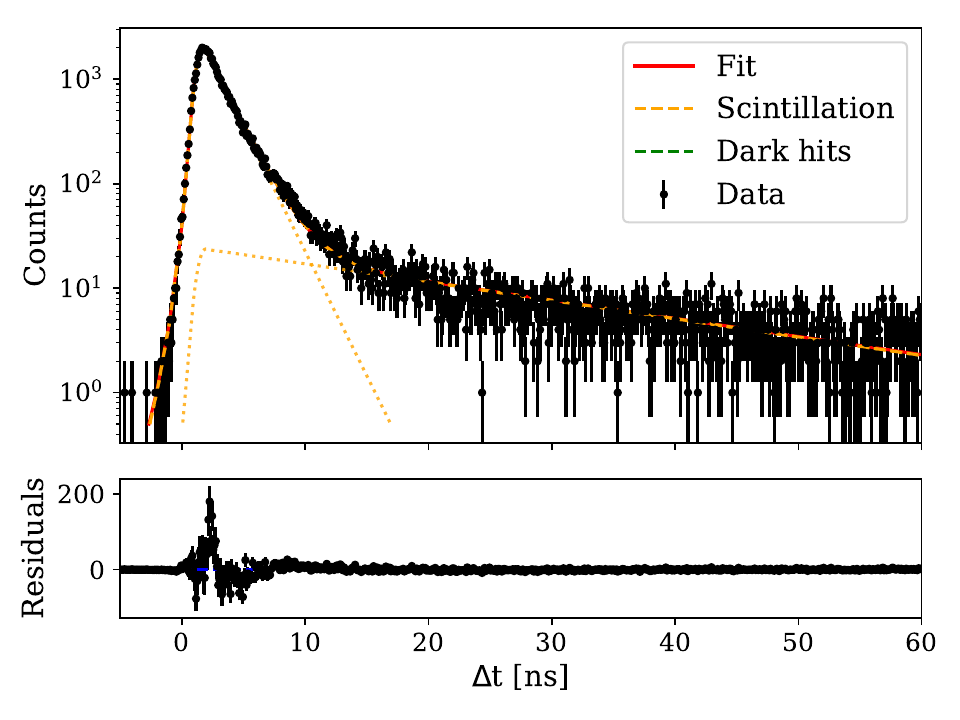}
    \includegraphics[width=0.330\textwidth]{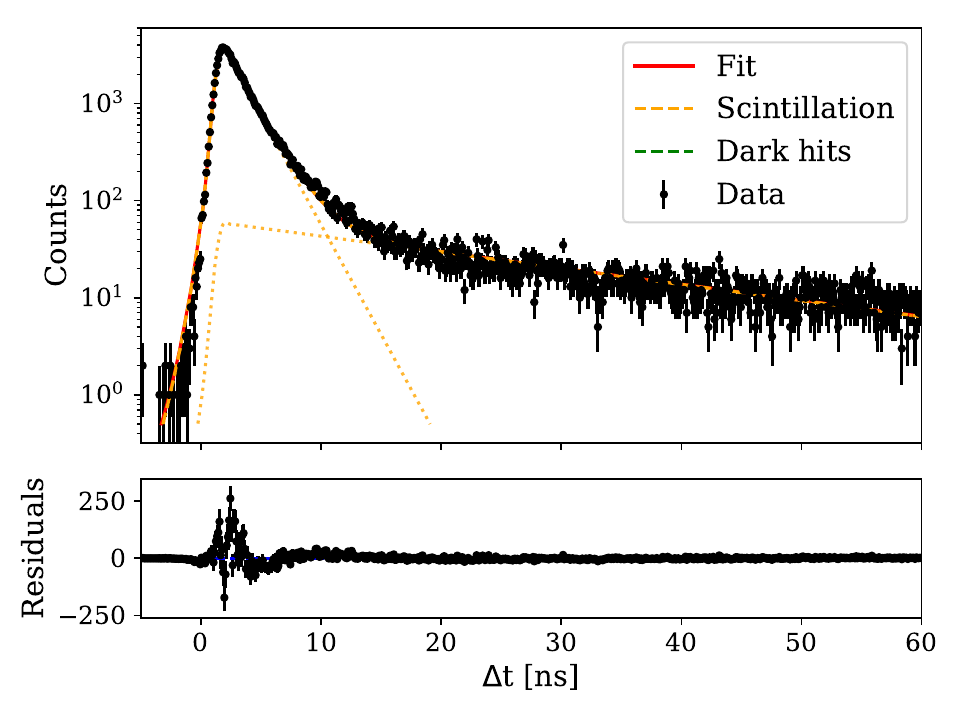}
    \includegraphics[width=0.330\textwidth]{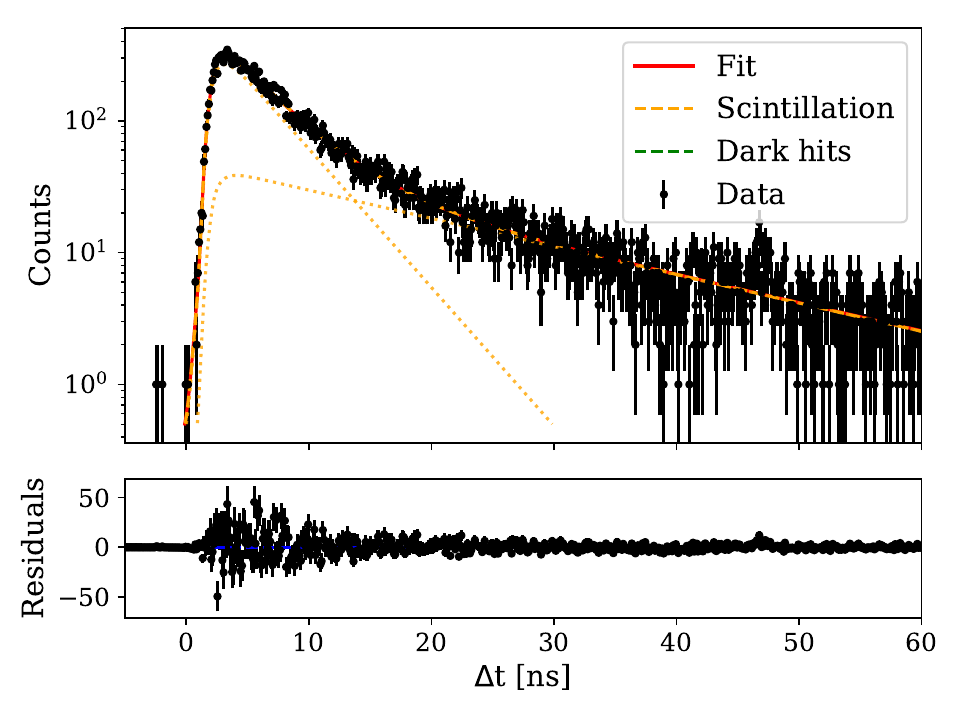}
    \caption{Distribution of detected time from trigger, with best-fit scintillation model overlaid, for 5\% \wbls{} (left), 10\% \wbls{} (middle), and LAB+PPO (right), in linear (top) and logarithmic (bottom) scales, using the $^{210}$Po $\alpha$ source. Dim dotted lines the show the individual scintillation decay modes.}
    \label{fig:timing_results}
\end{figure*}

    The timing data is shown, with best-fit scintillation time profile models, in Fig.~\ref{fig:timing_results}, and the associated parameters are summarized in Table~\ref{tbl:timing_results}. The fits are limited to an approximately 70~ns window, which does impact the total amount of late scintillation light that is included in the fit, particularly for the slower LAB+PPO sample. The mismodeling in the neighborhood of the peak, observable in the structure of the model residuals, can be associated with imperfect modeling of the trigger profile, which can be attributed to possible bias in the occupacy fits described in Sect.~\ref{sec:occupancy_spectrum}. Such bias can result from biased SPE calibration, of which there are numerous potential sources. One avenue for improvement is to fit for the SPE response simultaneous with the occupancy spectrum, using external calibraton data as a constraint, but such ventures are beyond the scope of this work. Mismodeling of the system response principally affects, of the scintillation parameters, the rise time, while the decay times are relatively robust. As timing-based PID performance is driven by the long-timescale decay behavior, the rise times listed in Table~\ref{tbl:timing_results} are deemed to have a negligible impact on the particle identification capabilities.

    In comparison to the scintillation timing from $\beta$ radiation of the \wbls{} mixtures, presented in Ref. \cite{Kaptanoglu:2021prv}, we find that the scintillation timing for $\alpha$ radiation is generally slower. This can be most clearly identified by comparing the value of $A_{1}$, which is around 95\% for the 5\% and 10\% \wbls{} for $\beta$ excitation. The smaller values of $A_{1}$ for $\alpha$ radiation indicate a higher fraction of the light is emitted later in time, causing the emission timing to be generally slower. This difference in timing can be leverage in particle identification techniques, presented in the following section.

\begin{table}[t!]
    \centering
	\caption{Best-fit time profile parameters for 5\% \wbls{}. 10\% \wbls{}, and LAB+PPO in response to $\alpha$ radiation. The time profile parameters for $\beta$ radiation can be found in Ref. \cite{Kaptanoglu:2021prv}.}
    \begin{tabular}{c | c | c | c} \hline \hline
        \- &
            5\% \wbls{} &
           10\% \wbls{} &
            LAB+PPO{} \\
        \hline
        $\tau_{R}\;[\SI{}{\pico\second}]$      &
            $169\plusminus{15}{15}$ &
            $129\plusminus{13}{13}$ &
            $709\plusminus{49}{49}$ \\
        $\tau_{1}\;[\SI{}{\nano\second}]$      &
            $1.82\plusminus{0.01}{0.01}$ &
            $1.92\plusminus{0.01}{0.01}$ &
            $4.13\plusminus{0.13}{0.13}$ \\
        $\tau_{2}\;[\SI{}{\nano\second}]$      &
            $24.7\plusminus{0.8}{0.8}$ &
            $26.1\plusminus{0.5}{0.5}$ &
            $20.3\plusminus{0.8}{0.8}$ \\
        $A_{1}\;[\%]$                   &
            $89.7\plusminus{0.2}{0.2}$ &
            $86.6\plusminus{0.1}{0.1}$ &
            $64.7\plusminus{1.3}{1.3}$ \\
        \hline
        $\chi^{2}$ &
            1430 &
            1685 &
            1316 \\
        $\ndf$ &
            1194 &
            1194 &
            1194 \\ \hline
    \end{tabular}
    \label{tbl:timing_results}
\end{table}

%% file: pid.tex
\section{Particle Identification Performance}\label{sec:pid}

    As discussed in Sect.~\ref{sec:intro}, signal and background processes for MeV-scale neutrino physics often arise from different particle species. For example, there is particular interest in the community in rejecting background $\alpha$s from radioactive decays, thereby improving the selection of electron-like events associated with neutrino interactions.

    The properties of each of the two main sources of photons in optical detectors, Cherenkov radiation and scintillation, can be leveraged to attempt separation of event samples by particle type. Cherenkov emission is prompt in time, is broad-spectrum in wavelength, is directional, and has a particle-mass-dependent energy threshold. Scintillation emission occurs over longer time scales, is typically confined to a relatively narrow waveband, is isotropic, and has effectively no energy threshold. However, the characteristics of the scintillation response are particle-dependent via so-called quenching mechanisms, for example as discussed in Sect.~\ref{sec:light_yield} and Sect.~\ref{sec:timing}. As such, Cherenkov and scintillation photons will present differently in timing, wavelength, and angular distributions, as well as in relative proportion as a function of energy, for different types of particles.

    This varied response of media to different particle species creates ample potential for PID capabilities. Traditional liquid scintillator detectors rely on timing-based separation mechanisms, but the added lever of finer knowledge of the distinct Cherenkov and scintillation signals has the potential to improve  discrimination power even more in \wbls{}. This is more pertinent at energies relevant to this study where $\alpha$s will be below Cherenkov threshold. While technologies are being investigated to harness the full potential of the time-based differences using ultra-fast photosensors and precise waveform digitization to allow for performance beyond the single-hit counting regime \cite{Kaptanoglu:2021prv}, chromatic differences using spectrally-sorting filters \cite{Kaptanoglu:2019gtg}, and angular differences using high-coverage detectors with sophisticated reconstruction algorithms \cite{Anderson:2022lbb}, we focus solely on extending the time-based PID analogy from liquid scintillator detectors to hybrid detectors deploying \wbls{} and fast-timing PMTs.

    In order to understand the impact of the timing and light yield measurements beyond the benchtop scale, we simulate realistic detector configurations of various sizes. We the assess the timing-based PID capabilities thereof to understand the level of background rejection, based on particle type, achievable with \wbls{}.

\subsection{Simulations}\label{sec:sim}
We study detectors at the \SI{1}{~\tonne}{}, \SI{1}{~\kilo\tonne}{} and \SI{100}{~\kilo\tonne}{} scales with the following configurations:

\begin{enumerate}
    \item A $\sim$\SI{4}{~\tonne{}} detector of \eos{}-like geometry \cite{Anderson:2022lbb}, primarily employing Hamamatsu 8-\inch{} R14688-100 PMTs \cite{r14688} with some additional models for a total number of 231 PMTs, yielding a photocoverage of $\sim$40\%
    \item A $\sim \SI{1}{~\kilo\tonne}$ right cylindrical detector with \SI{5.4}{~\meter} fiducial radius and $\sim 54\%$ photocoverage using $\sim 3,000$ 12-\inch{} PMTs with equivalent quantum efficiency and time response to the R14688-100 %3046
    \item A $\sim \SI{100}{~\kilo\tonne}$ right cylindrical detector with \SI{25.2}{~\meter} fiducial radius and $\sim 85\%$ photocoverage, via $\sim 47,000$ PMTs of the same hypothetical model as the $\SI{1}{~\kilo\tonne}$ configuration %47086
\end{enumerate}
The detector volumes were determined assuming material density equivalent to water (i.e. \SI{1}{~\gram{\per\centi\meter}}$^{3}$) for the given mass.

    We simulate each detector configuration filled with 5\% \wbls{}, 10\% \wbls{}, and LAB+PPO. The scintillation time profiles and light yields of \wbls{} are taken from this work or previous measurements \cite{Caravaca:2020lfs,Kaptanoglu:2021prv}, as appropriate. For LAB+PPO, the time profile under $\alpha$ excitation is used as measured in this work, whereas the time profile under $\beta$ excitation is obtained using similar methodology to \cite{Kaptanoglu:2021prv} and the light yield parameters are obtained from similar methodology to \cite{Caravaca:2020lfs} and building on work from the SNO+ collaboration \cite{SNO:2020fhu}. While Sect.~\ref{sec:light_yield_results_alphas} reports average photon production for $\alpha$ excitation, as opposed to specific model parameters, simulation of large scale detectors requires a definite choice of parameter values. Here we choose parameters such that the scintillation efficiency $S$ for electron and $\alpha$ excitation are both equal to the value measured for electrons. These values are listed in Table~\ref{tab:lypars}.

\begin{table*}[t!]
    \centering
    \caption{The chosen light yield parameters for the three materials of interest used in the particle identification simulations. The scintillation efficiency $S$ is chosen to be the same for both species and is thus listed only once.}
    \begin{tabular}{c|c|c|c} \hline \hline
        Material & $S$ [ph\SI{}{{\per\MeV}}]& $\kB$ for $\beta$s [\SI{}{\milli\meter{\per\MeV}}] & $\kB$ for $\alpha$s [\SI{}{\milli\meter{\per\MeV}}] \\ \hline
          5\% \wbls{} & 754.0 & 0.074 & 0.092 \\
          10\% \wbls{} & 1380.0 & 0.074  & 0.088 \\
         LAB+PPO & 12200 & 0.074 & 0.076 \\
         \hline
    \end{tabular}
    \label{tab:lypars}
\end{table*}

    We focus on determination of PID performance at the energy of the \SI{5.3}{~\MeV} $\alpha$ from the \isotope{Po}{210} decay, as this is commonly the most prevalent $\alpha$ background from radioactive contaminants in liquid scintillator experiments. To do so, we simulate the \isotope{Po}{210} $\alpha$ decay and monoenergetic $\beta$s with kinetic energies that give an equivalent number of optical photons produced as the \isotope{Po}{210} decays. The corresponding $\beta$ energy varies by material due to differences in quenching, and is shown in Table~\ref{tab:betaen}. Several other decays of interest, such as \isotope{Po}{212} and \isotope{Po}{214} occur at higher energies, and thus the performance determined in this section should be conservative for those decays, as the added light will improve the performance as discussed below. All events are simulated at the center of the detector configurations and isotropically in direction.

\begin{table}[t!]
    \centering
	\caption{The kinetic energy of $\beta$s corresponding to an equivalent number of optical photons produced as for \isotope{Po}{210} $\alpha$ decays in each material.}
    \begin{tabular}{c|c} \hline \hline
          Material & Equivalent $\beta$ energy [\SI{}{\keV}] \\ \hline
          5\% \wbls{} & 395 \\
          10\% \wbls{} & 401 \\
         LAB+PPO & 501 \\
         \hline
    \end{tabular}
    \label{tab:betaen}
\end{table}

\subsection{Classification Routine} \label{sec:pid_routine}

    To study the PID capabilities for these materials, we employ a likelihood-ratio statistic calculated using the hit time residuals for each event. The hit time residual is defined as
\begin{equation}\label{eq:tres}
t_\text{res} = t_\text{hit} - t_\text{tof} - t_\text{vertex},
\end{equation}
where $t_\text{hit}$ is the hit time for the PE as recorded by the PMT (with time response included), $t_\text{tof}$ is the estimated time-of-flight from event vertex to PMT and $t_\text{vertex}$ is the time of the event vertex. For the time-of-flight calculation, in all cases we assume straight line paths from the event vertex to a hit PMT in the target medium, i.e. no refractions or reflections at interfaces are considered, nor is the effect of scattered light, though these processes are present in the simulation. Some optical effects, e.g. scattering, introduce a common smearing which degrades performance, but the assumption of unrefracted paths does not affect the results, as likelihood ratios are invariant under differentiable transformations. While the index of refraction is varied appropriately by material, we assume a $\SI{400}{~\nano\meter}$ wavelength for all photons when calculating the time-of-flight. This wavelength is near the peaks of both the R14688 quantum efficiency and the scintillation emission spectrum of LAB+PPO and \wbls{} \cite{r14688}\cite{Bonventre:2018hyd}\cite{Land:2020oiz}. At 400 nm the modeled absorption and scattering lengths of each of the \wbls{} materials are longer than $\SI{10}{~\meter}$.

    We define a classifier value $c$ for an event as the average likelihood-ratio of the observed photon times, comparing the $\alpha$- and $\beta$-hypotheses. That is,
\begin{equation}\label{eq:class}
c = \frac{1}{N}\sum_{i=0}^{N} \ln P(t_i|\alpha) - \ln P(t_i|\beta),
\end{equation}
where the sum is taken over the $N$ distinct photons detected, which, depending on implementation details for a real detector, is generally either the collection of PMTs which were ``hit,'' or the collection of individual photoelectrons detected, subject to cuts on the timing of the hits and other factors. $P(t_i | \alpha)$ is the probability of observing time residual $t_i$ given the event is an $\alpha$, and $P(t_i | \beta)$ is defined analogously for $\beta$s. $P(t | \alpha)$ and $P(t | \beta)$ are probability density functions (PDFs) determined from simulation, as discussed above. Dedicated PDFs are generated for each material and detector configuration. As the classifier is defined as a sample average, the classifier distributions can be understood via the central limit theorem: the mean classifier value for each particle is the likelihood-ratio averaged over the time-residual distribution for that particle, and the distribution is asymptotically Gaussian with standard deviation inversely proportional to $\sqrt{N}$. This naturally facilitates the comparison of different materials and detector configurations, as it decouples the intrinsic classification power (the mean values), driven by the different emission time profiles, from that achievable in any particular deployment, determined by the amount of light collected.

We generate time-residual PDFs for two extreme photon-counting scenarios: simple ``hits,'' wherein only the time of the first photon incident on a PMT is known, and full photoelectron disambiguation, in which the time of every photoelectron is known, regardless of per-PMT pileup. The scintillation emission time profiles measured in this work were measured over a ($\sim \SI{60}{~\nano\second}$) time scale, whereas the event window length in real detectors may be much longer, on the order of $\sim \SI{400}{~\nano\second}$. To compromise between these two time scales, we use an analysis window of $\SI{-10}{~\nano\second}$ to $\SI{220}{~\nano\second}$. This allows for fair inclusion of optical effects, such as Rayleigh scattering, without extending into a regime where unmeasured scintillation decay modes may dominate the detector response. Vertex reconstruction is not performed, and instead we use the true position and time of each event in place of the reconstructed vertex, though the effect of simply smearing this vertex by a characteristic resolution is studied. The emission timing measurements performed in the 60~ns window are used to extrapolate into the longer time-window, which is determined to under predict the scintillation timing tail for LAB+PPO in comparison to work that used longer analysis windows~\cite{SNO:2020fhu,LOMBARDI2013133}. This has less of an impact on the WbLS mixtures, which are significantly faster and emit less light beyond the 60~ns analysis window. 

\begin{figure*}[t!]
    \centering
    \includegraphics[width=0.65\columnwidth]{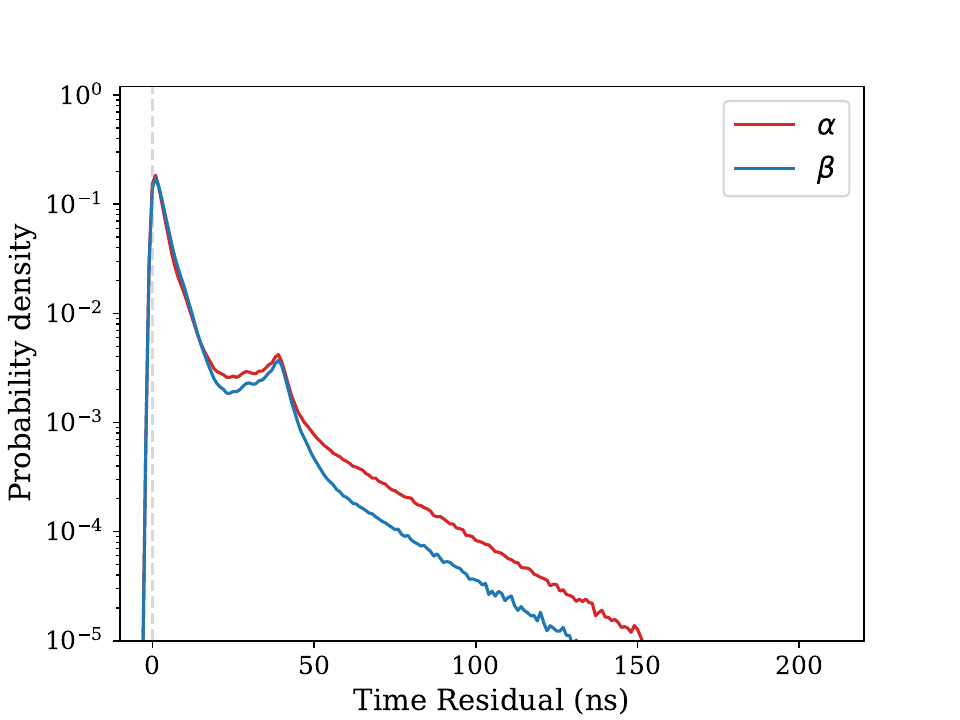}
    \includegraphics[width=0.65\columnwidth]{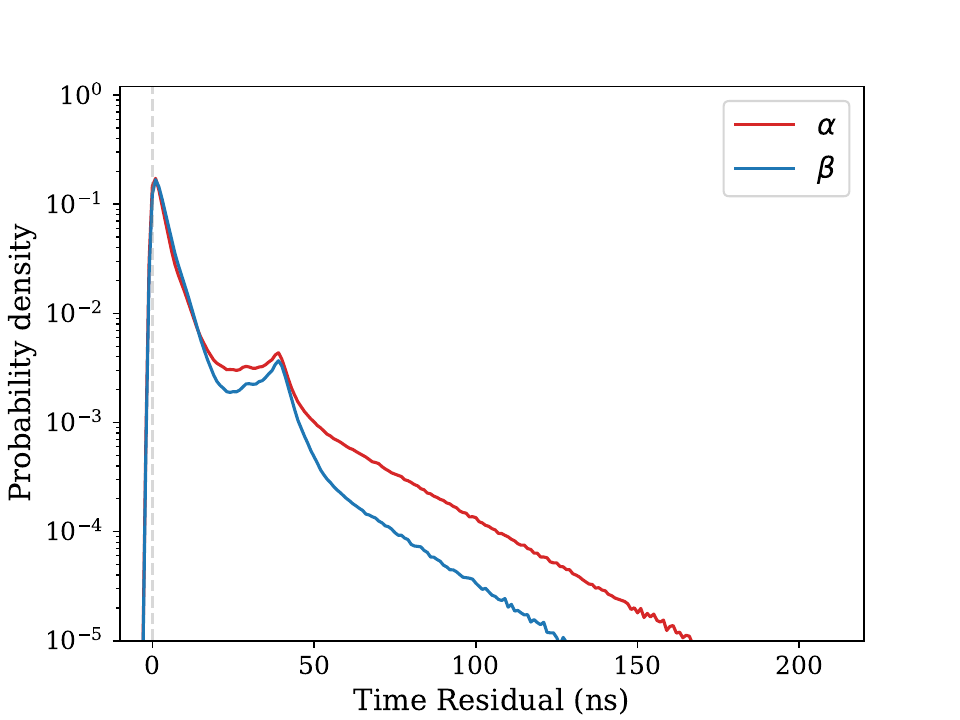}
    \includegraphics[width=0.65\columnwidth]{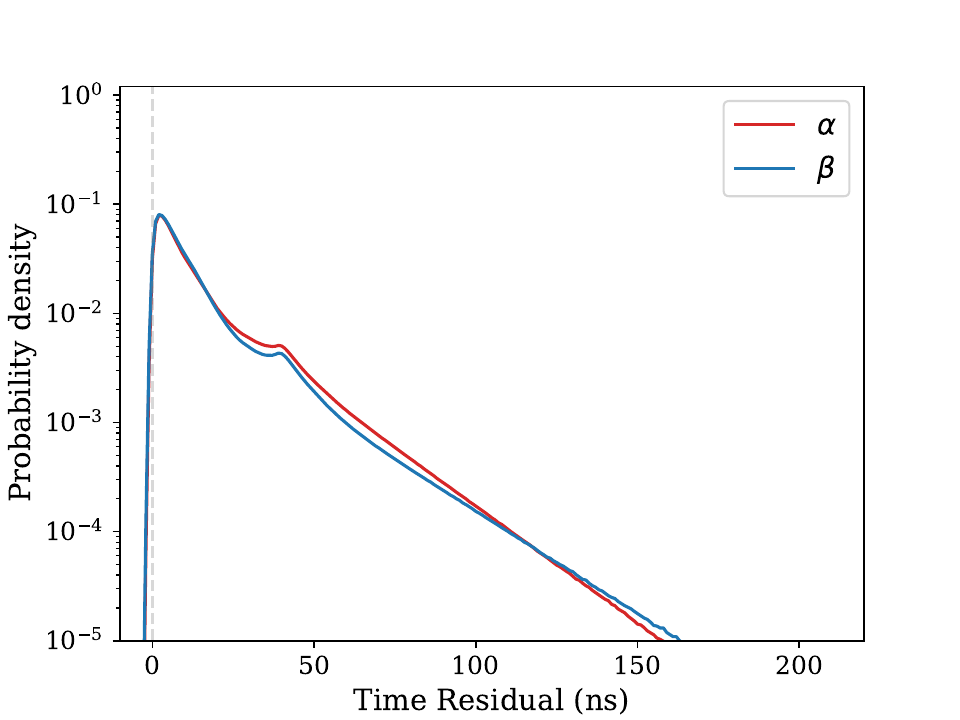}
	\caption{Comparisons of the $\alpha$ and $\beta$ time residuals, using all detected PEs, in the \eos{}-like detector for 5\% \wbls{} (left), 10\% \wbls{} (center), and LAB+PPO (right). All events are simulated at the center of the detector, uniformly in direction. The $\alpha$ particles are produced from $^{210}$Po decays and the $\beta$ particles are simulated with energies from Table \ref{tab:betaen}.}
    \label{fig:pidpdfs}
\end{figure*}

\begin{figure*}[t!]
    \centering
    \includegraphics[width=0.65\columnwidth]{pid/pidreviewplot_Eos_wbls_10pct_allpe_area.pdf}
    \includegraphics[width=0.65\columnwidth]{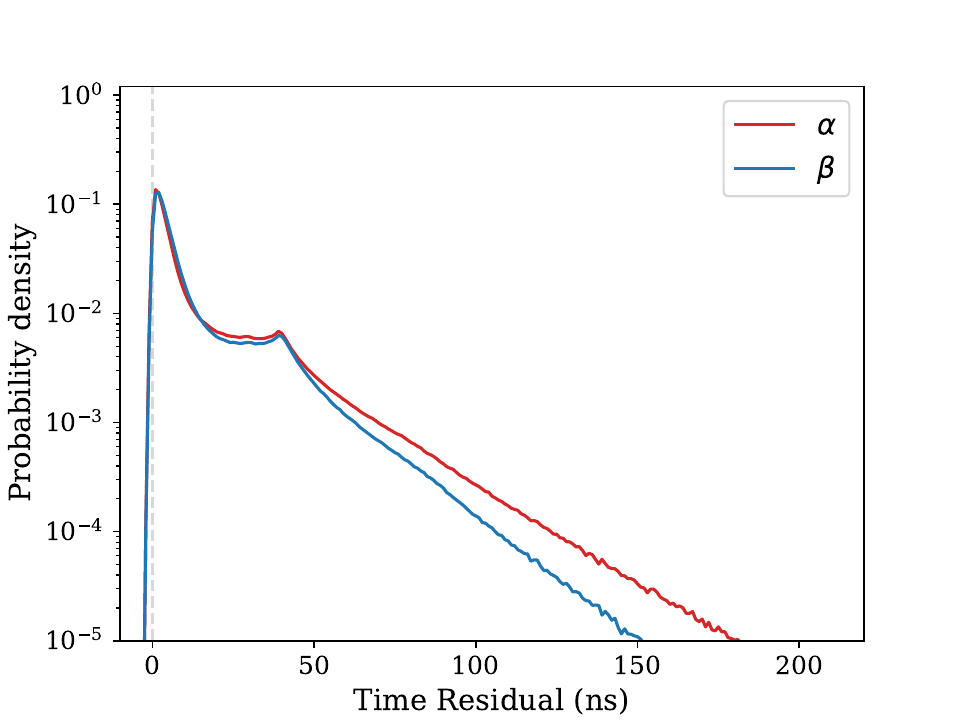}
    \includegraphics[width=0.65\columnwidth]{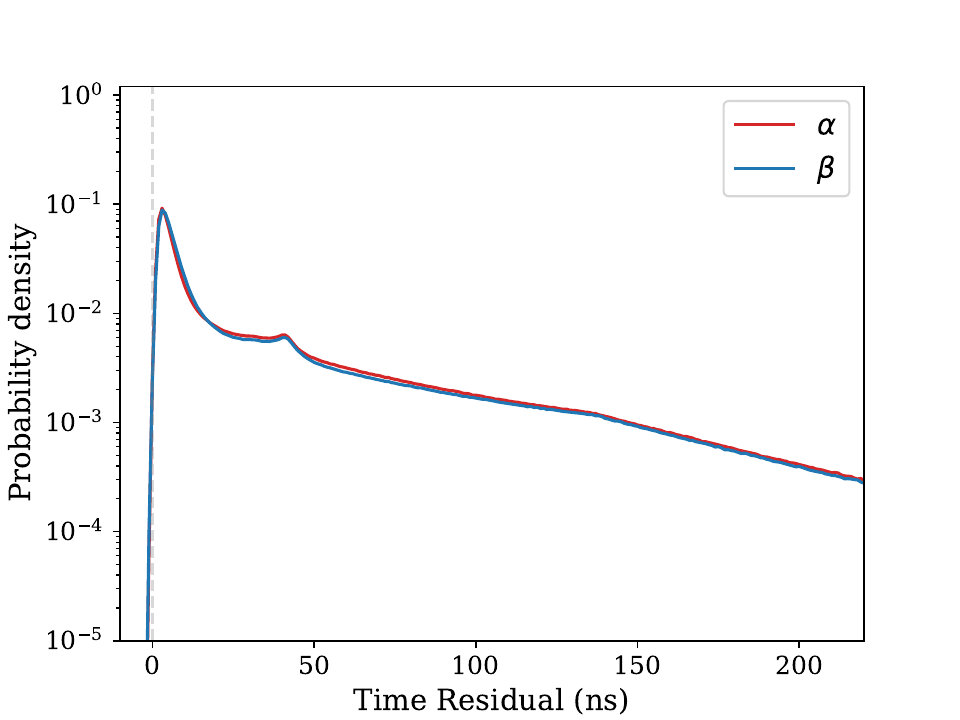}
        \caption{Comparisons of the $\alpha$ and $\beta$ time residuals, using all detected PEs, in an Eos-like detector (left), a \SI{1}{\kilo\tonne{}} detector with 54\%~photocoverage (center), and a \SI{100}{\kilo\tonne{}} detector with 85\%~photocoverage (right) for 10\% \wbls{}. The $\alpha$ particles are produced from $^{210}$Po decays and the $\beta$ particles are simulated with energies from Table \ref{tab:betaen}.}
    \label{fig:pidpdfsmultikt}
\end{figure*}

\begin{figure}[t!]
    \centering
	\includegraphics[width=0.8\columnwidth]{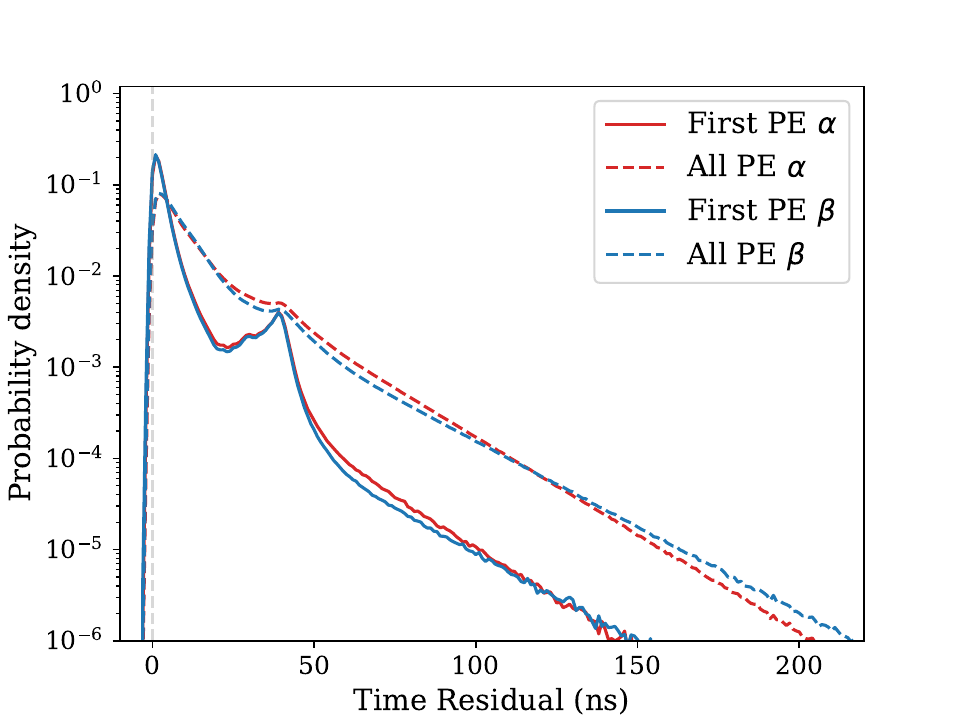}
    \caption{Comparison of the ``first PE'' (solid) and ``all PE'' (dashed) PDFs for both species in LAB+PPO in the \eos{}-like detector. This detector configuration is heavily multi-PE, resulting in a significant difference in the shape of the PDFs observed in the two cases. All events are simulated at the center of the detector, uniformly in direction. The $\alpha$ particles are produced from $^{210}$Po decays and the $\beta$ particles are simulated with energies from Table \ref{tab:betaen}.}
    \label{fig:peoverlay}
\end{figure}

Example of time-residual PDFs can be found in Fig.~\ref{fig:pidpdfs}, which shows the time-residuals for each different material in \eos{}, and in Fig.~\ref{fig:pidpdfsmultikt}, which shows the time-residuals for each different detector size with 10\% \wbls{}. The additional peak around \SI{40}{~\nano\second} is caused by PMT late pulsing, typical of large area PMTs. The \eos{}-like detector, owing to its relatively small size, observes a relatively high proportion of multi-PE PMT hits, which is distinct from the larger detectors which operate largely in the SPE regime. As such, the difference between ``first PE'' and ``all PE'' PDFs, and hence the corresponding PID performance, in the \eos{}-like detector is larger than for the other two detectors. In particular for LAB+PPO, the ``first PE'' PDFs in the \eos{}-like detector are very similar between the two species, resulting in poorer PID performance than would naively be expected for LAB+PPO. This comes as a result of all the PMTs in the detector registering multiple PE, and so the ``first PE'' PDFs lack a substantial amount of information, particularly from hits coming at later times, as shown in Fig.~\ref{fig:peoverlay}. With modern detectors, such as \eos{}, seeking to differentiate hits at least in the ``few PE'' regime using photosensors with faster timing and better readout schemes, we present results based on PDFs using all hit times. 

\subsection{Analysis Methods}\label{sec:pidanalysis}

    The classifier value is a quantity for which, ideally, the distributions associated with $\alpha$s and $\beta$s have little overlap, which would enable an efficient selection cut to be employed in the course of a physics analysis. As positive values are associated with $\alpha$-like events, and negative values with $\beta$-like events, a simple threshold value can be used to perform classification. We study these distributions and the resulting classification performance both analytically, and using further simulation.

    As described in Sect.~\ref{sec:pid_routine}, these distributions are asymptotically Gaussian, and as such can summarized by their means and standard deviations. Mathematically, this translates to computing the first and second moments of the log-likelihood ratio: the former is the mean classifier value, and the latter, when scaled by $\sqrt{N}$, is the standard deviation. We calculate these values for each detector configuration via numerical integration, to offer the full classifier distributions (at least, asymptotically) in a convenient form, to facilitate follow-on performance studies

\begin{figure*}[t!]
\centering
    \includegraphics[width=0.65\columnwidth]{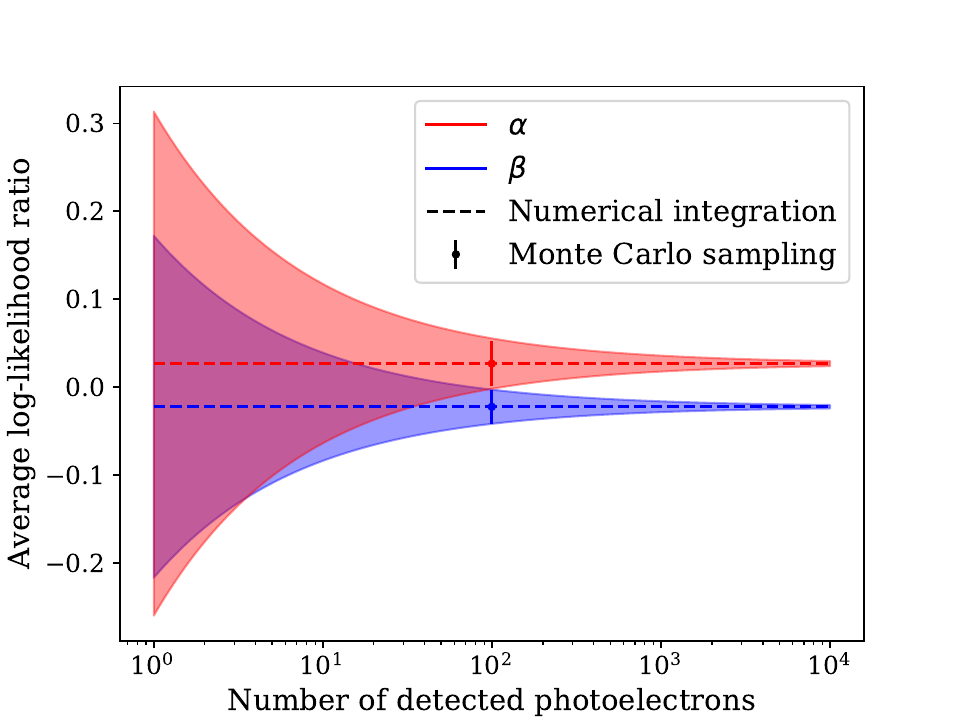}
    \includegraphics[width=0.65\columnwidth]{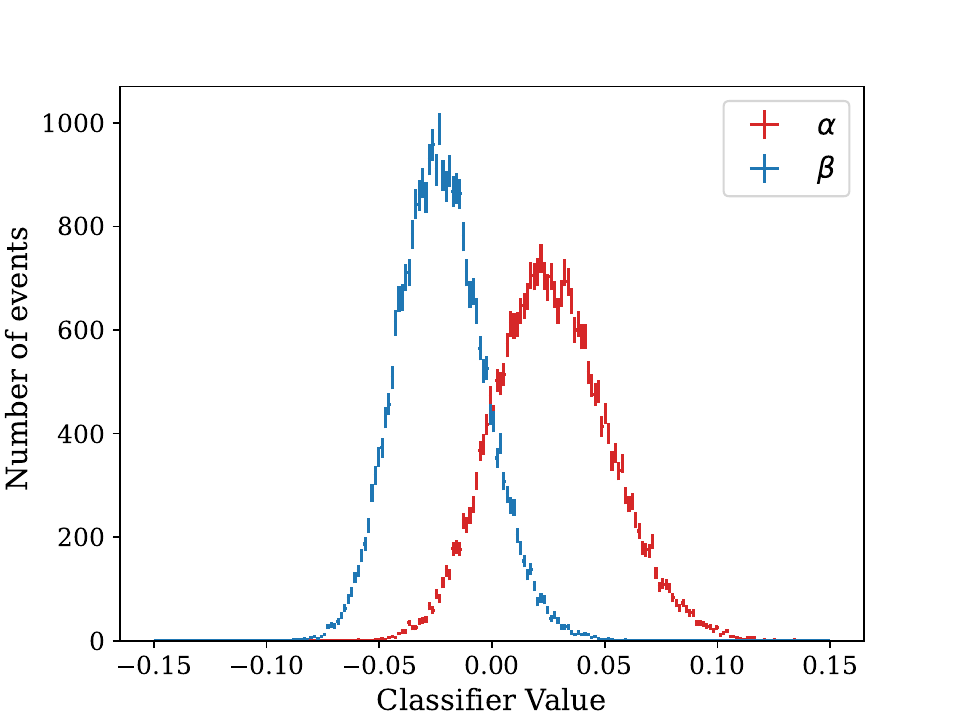}
    \includegraphics[width=0.65\columnwidth]{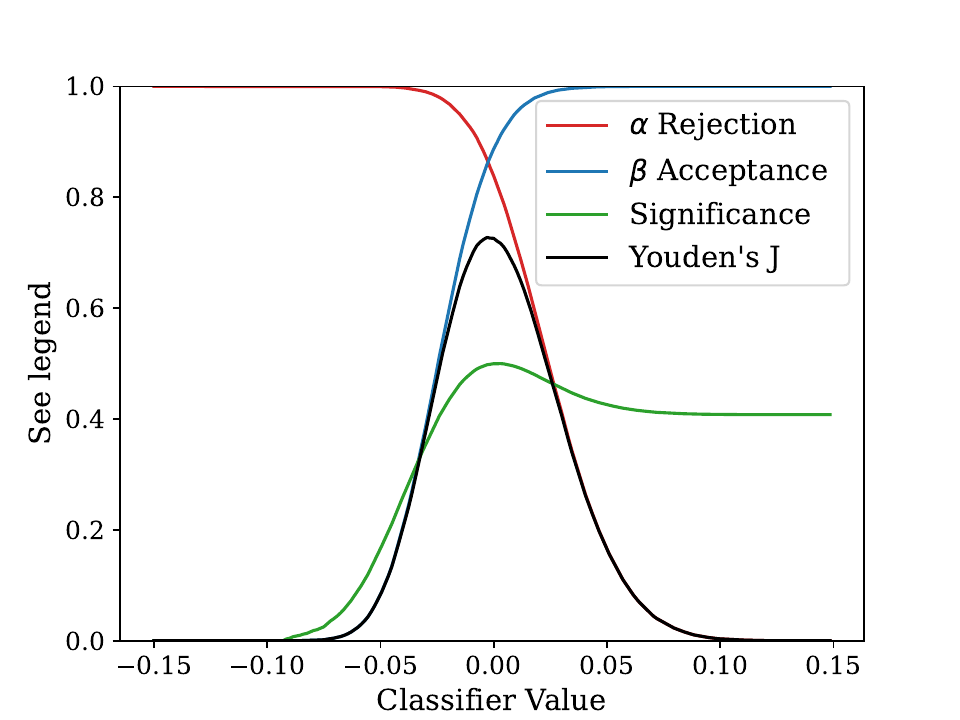}
	\caption{Dependence of classifier distributions on the number of photoelectrons detected for $\alpha$ (red) and $\beta$ (blue) particles (left), nominal classifier distributions for $\alpha$ and $\beta$ particles (center), and nominal PID performance figures-of-merit as a function of cut value (right), for 10\% \wbls{} in an \eos{}-like detector. The shaded bands in the left plot denote the standard deviation of the distributions predicted by numerical integration, and the data points denote the results of explicit Monte Carlo sampling, i.e. the distributions shown in the center plot.}
    \label{fig:wbls10pctallpenormresults}
\end{figure*}

    Then, using further full MC simulations to sample from the non-asymptotic classifier distributions, we study various figures of merit which quantify the classification performance, as a function of threshold value. An example pair of classifier distributions, both sampled from simulation and expressed as a Gaussian, can be found in Fig.~\ref{fig:wbls10pctallpenormresults}, which corresponds to a 10\% \wbls{}-filled \eos{}-like detector. The two approaches are compared in the leftmost figure, which shows the compatibility of the full MC sampling and asymptotic distributions, indicating that the non-Gaussian components of the classifier distributions at this light level are small.

    As electrons are usually considered a ``signal'' in analysis of neutrino detector data, we label $\beta$ events as signal and $\alpha$ events as background in this work. The specific classification figures of merit considered are:
\begin{enumerate}
    \item Sample purity: $N_\beta(\ccut)/N_\text{tot}(\ccut)$
    \item Signal acceptance: $N_\beta(\ccut)/N_{\beta, \text{tot}}$
    \item Background rejection: $N_\alpha(\ccut)/N_{\alpha, \text{tot}}$
    \item Significance: $N_\beta(\ccut)/\sqrt{N_\beta(\ccut) + N_\alpha(\ccut)}$
    \item Youden's J: $N_\beta(\ccut)/N_{\beta, \text{tot}}- N_\alpha(\ccut)/N_{\alpha, \text{tot}}$
\end{enumerate}
where $\ccut$ is the cut value on classifier value; $N_{\alpha}\pp{\ccut}$ and $N_{\beta}\pp{\ccut}$ are the numbers of $\alpha$ and $\beta$ events selected from the sample, respectively; $N_{\alpha, \text{tot}}$ and $N_{\beta, \text{tot}}$ are the total numbers of $\alpha$ and $\beta$ events in the sample, respectively; and $N_{\text{tot}}\pp{\ccut} = N_{\alpha}\pp{\ccut} + N_{\beta}\pp{\ccut}$ is the total number of selected events. The selection is performed such that all events with classifier value less than or equal to $\ccut$ are considered to pass the selection cut, and all those with classifier value greater than $\ccut$ fail the selection cut. Each of these quantities is computed as function of the cut value $\ccut$.

\subsection{Results}\label{sec:pidres}

    The first and second moments of the classifier distributions, i.e. the means and single-sample standard deviations of the log-likelihood ratio, as well as the mean numbers of detected photoelectrons, are listed in Table~\ref{tbl:simfree_pid_alpha}, Table~\ref{tbl:simfree_pid_beta}, and Table~\ref{tbl:sim_npe}. The PID figures of merit for a 10\% \wbls{}-filled \eos{}-like detector are shown as a function of the classifier cut value in the right panel of Fig.~\ref{fig:wbls10pctallpenormresults}, and Table~\ref{tab:eosresults} lists the background rejections associated with a 90\% signal acceptance for \eos{}-like detectors of various target media. From Table~\ref{tab:eosresults}, we see that the $\alpha$ rejection improves with higher light yields from the increasing scintillator concentration. 

\begin{table*}[t!]
    \centering
    \caption{Mean and single-sample standard deviation of classifier values for the \isotope{Po}{210} simulations in the various detectors.}
    \begin{tabular}{r | c | c | c || c | c | c} \hline \hline
        \- &
            \multicolumn{3}{c||}{Mean} &
            \multicolumn{3}{c}{Standard deviation} \\
        \- &
            5\% \wbls{} &
            10\% \wbls{} &
            LAB+PPO &
            5\% \wbls{} &
            10\% \wbls{} &
            LAB+PPO \\
        \hline
        \SI{4}{~\tonne{}} &
            $1.11 \times 10^{-2}$ &
            $2.68 \times 10^{-2}$ &
            $4.42 \times 10^{-3}$ &
            $1.77 \times 10^{-1}$ &
            $2.87 \times 10^{-1}$ &
            $9.58 \times 10^{-2}$ \\
        \SI{1}{~\tonne{}} &
            $5.67 \times 10^{-3}$ &
            $1.32 \times 10^{-2}$ &
            $2.15 \times 10^{-3}$ &
            $1.22 \times 10^{-1}$ &
            $1.68 \times 10^{-1}$ &
            $6.47 \times 10^{-2}$ \\
        \SI{100}{~\kilo\tonne{}} &
            $2.70 \times 10^{-3}$ &
            $5.38 \times 10^{-3}$ &
            $1.12 \times 10^{-3}$ &
            $7.27 \times 10^{-2}$ &
            $1.03 \times 10^{-1}$ &
            $4.26 \times 10^{-2}$ \\
        \hline
    \end{tabular}
    \label{tbl:simfree_pid_alpha}
\end{table*}

\begin{table*}[t!]
    \centering
	\caption{Mean and single-sample standard deviation of classifier values for the $\beta$ excitation (energies from Table \ref{tab:betaen}) in the various detectors.}
    \begin{tabular}{c | c | c | c || c | c | c} \hline \hline
        \- &
            \multicolumn{3}{c||}{Mean} &
            \multicolumn{3}{c}{Standard deviation} \\
        \- &
            5\% \wbls{} &
            10\% \wbls{} &
            LAB+PPO &
            5\% \wbls{} &
            10\% \wbls{} &
            LAB+PPO \\
        \hline
        \SI{4}{~\tonne{}} &
            $-1.00 \times 10^{-2}$ &
            $-2.23 \times 10^{-2}$ &
            $-4.27 \times 10^{-3}$ &
            $1.36 \times 10^{-1}$ &
            $1.94 \times 10^{-1}$ &
            $9.08 \times 10^{-2}$ \\
        \SI{1}{~\tonne{}} &
            $-5.53 \times 10^{-3}$ &
            $-1.24 \times 10^{-2}$ &
            $-2.25 \times 10^{-3}$ &
            $1.04 \times 10^{-1}$ &
            $1.53 \times 10^{-1}$ &
            $7.09 \times 10^{-2}$ \\
        \SI{100}{~\kilo\tonne{}} &
            $-2.76 \times 10^{-3}$ &
            $-5.45 \times 10^{-3}$ &
            $-1.57 \times 10^{-3}$ &
            $7.51 \times 10^{-2}$ &
            $1.05 \times 10^{-1}$ &
            $8.03 \times 10^{-2}$ \\
        \hline
    \end{tabular}
    \label{tbl:simfree_pid_beta}
\end{table*}

\begin{table}[t!]
    \centering
	\caption{The mean detected PE for each detector configuration for the simulated \isotope{Po}{210}. Uncertainties are less than 1 PE in all cases.}
    \begin{tabular}{r | c | r}
        \hline
        \hline
        Nominal mass             & Material      &   Mean $N_{\text{PE}}$ \\
        \hline
        \SI{4}{~\tonne{}}        & 5\% \wbls{}   &    52.6 \\
        \SI{1}{~\kilo\tonne{}}   & 5\% \wbls{}   &    57.5 \\
        \SI{100}{~\kilo\tonne{}} & 5\% \wbls{}   &    77.6 \\
        \hline
        \SI{4}{~\tonne{}}        & 10\% \wbls{}  &    99.7 \\
        \SI{1}{~\kilo\tonne{}}   & 10\% \wbls{}  &   108.5 \\
        \SI{100}{~\kilo\tonne{}} & 10\% \wbls{}  &   145.6 \\
        \hline
        \SI{4}{~\tonne{}}        & LAB+PPO       &  1039.6 \\
        \SI{1}{~\kilo\tonne{}}   & LAB+PPO       &   985.2 \\
        \SI{100}{~\kilo\tonne{}} & LAB+PPO       &  1039.7 \\
        \hline
    \end{tabular}
    \label{tbl:sim_npe}
\end{table}

    Vertex reconstruction resolution is generally robust across different detector sizes \cite{Land:2020oiz}, and hence the effect of vertex reconstruction error is quantified using a 5\%~\wbls{}-filled \eos{}-like detector, which, owing to its small size, is the most sensitive to the impact of resolution. Applying a \SI{10}{~\centi\meter{}} Gaussian smearing to each Cartesian coordinate of the true vertex position, representative of the achievable position resolution \cite{Land:2020oiz}, results in a nonzero but mild loss in performance of less than 0.5\% absolute background rejection. The degradation is smaller in larger detectors, where longer nominal photon times-of-flight dominate the fixed uncertainty introduced by fixed vertex resolution. This study does not account for possible complicated reconstruction features, such as tails in the vertex reconstruction, which could impact the PID.

\begin{table}[t!]
\centering
	\caption{$\alpha$ rejection in \eos{}-like for various materials, for cut values that yield a $\beta$ acceptance of 90\%. These results use the simulations performed with $^{210}$Po and the $\beta$ particles at energies from Table \ref{tab:betaen}.}
\begin{tabular}{c|c|c} \hline \hline
 Material      &   Cut Value &   $\alpha$ Rej. [\%]  \\
\hline
    5\% \wbls{}  &       0.013 &                      43.1  \\
    10\% \wbls{} &       0.002 &                      83.2  \\
    LAB+PPO      &      -0.001 &                      96.3  \\
\hline
\end{tabular}
\label{tab:eosresults}
\end{table}

\begin{figure*}[t!]
    \centering
    \includegraphics[width=0.65\columnwidth]{pid/pidreview_Eos_allpe_wbls_10pct_multi_nhit0to1000_BerkeleyAlphaBetaNorm_MPL.pdf}
    \includegraphics[width=0.65\columnwidth]{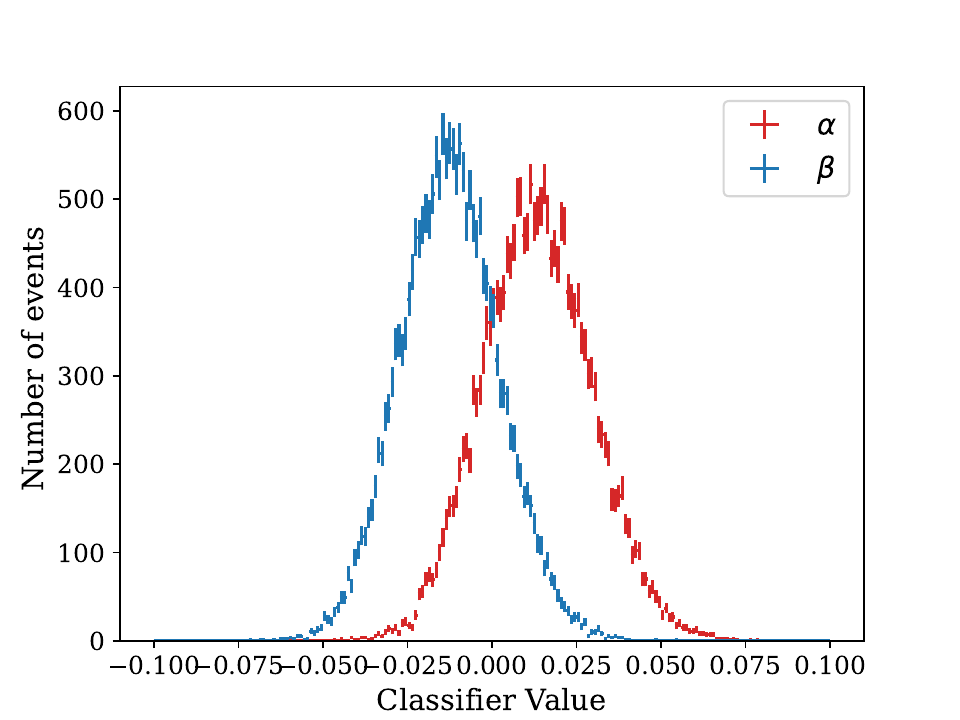}
    \includegraphics[width=0.65\columnwidth]{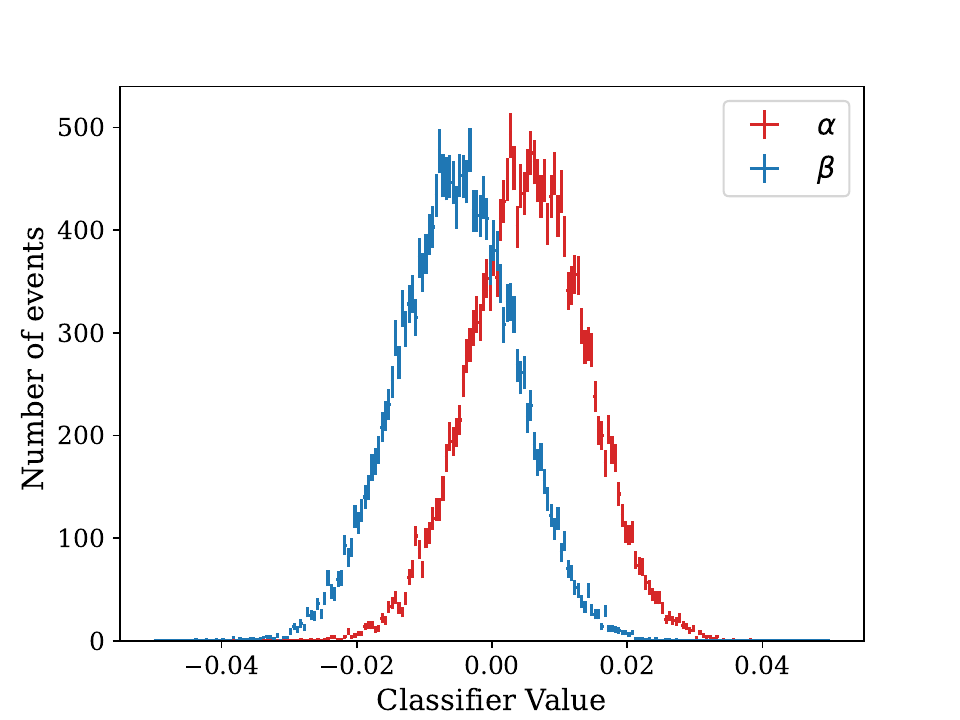}
    \caption{Comparisons of the classifier results in an Eos-like detector (left), a \SI{1}{\kilo\tonne{}} detector with 54\%~photocoverage (center), and a \SI{100}{\kilo\tonne{}} detector with 85\%~photocoverage (right) for 10\% \wbls.}
    \label{fig:pidresultsmultikt}
\end{figure*}

\begin{figure}[t!]
    \centering
	\includegraphics[width=0.8\columnwidth]{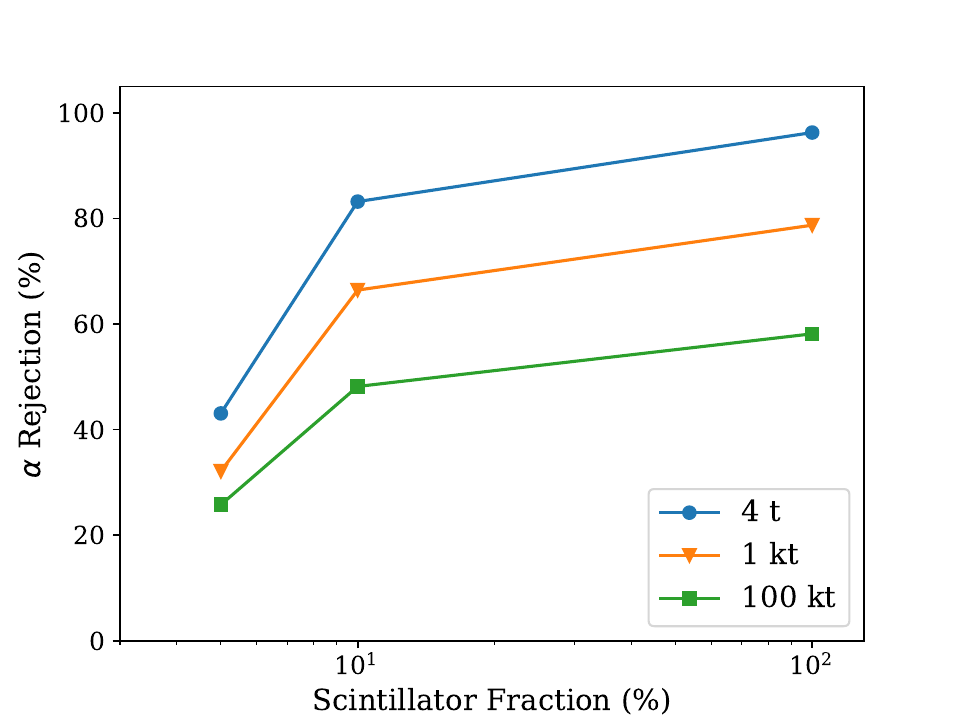}
	\caption{Comparison of the $\alpha$ rejection as a function of material scintillator fraction for the three detector concepts, assessed at the 90\% $\beta$ acceptance cut threshold. Notably, as discussed in Section \ref{sec:pidres}, using the LAB+PPO emission timing measured in~\cite{SNO:2020fhu} results in 100\% separation for all three detector configurations.} 
    \label{fig:rejsummaryplot}
\end{figure}

    The classifier distributions for 10\% \wbls{} for the three detectors are shown in Fig.~\ref{fig:pidresultsmultikt}. In all cases, the classifier distributions are visibly well-separated around the log-likelihood ratio of 0, with polarity as expected from construction. Table~\ref{tab:theiaresults} similarly reports the figures of merit associated with a signal acceptance of 90\%. The same trend of increasing performance with increasing scintillator fraction is observed in these detector configurations and is summarized in Fig.~\ref{fig:rejsummaryplot}, which shows the $\alpha$ rejection at 90\% $\beta$ acceptance for all three detectors as a function of scintillator fraction of the material.

\begin{table}[t!]
\centering
\caption{$\alpha$ rejection in \SI{1}{\kilo\tonne{}} and \SI{100}{\kilo\tonne{}} detectors for various materials, for cut values that yield $\beta$ acceptances of 90\%. These results use the simulations performed with $^{210}$Po and the $\beta$ particles at energies from Table \ref{tab:betaen}.}
\begin{tabular}{c|c|c|c} \hline
\hline
    Det.     & Material   &   Cut Value &   $\alpha$ Rej. [\%] \\
\hline
\SI{1}{~\kilo\tonne{}}   & 5\% \wbls{}  &       0.012 &                      32.1   \\
 \SI{1}{~\kilo\tonne{}}   & 10\% \wbls{} &       0.006 &                      66.4  \\
 \SI{1}{~\kilo\tonne{}}   & LAB+PPO      &       0.001 &                      78.7  \\
 \hline
 \SI{100}{~\kilo\tonne{}} & 5\% \wbls{}  &       0.008 &                      25.8  \\
 \SI{100}{~\kilo\tonne{}} & 10\% \wbls{} &       0.006 &                      48.2  \\
 \SI{100}{~\kilo\tonne{}} & LAB+PPO      &       0.001 &                      58.1  \\
\hline
\end{tabular}
\label{tab:theiaresults}
\end{table}

\begin{figure*}[t!]
    \centering
    \includegraphics[width=0.65\columnwidth]{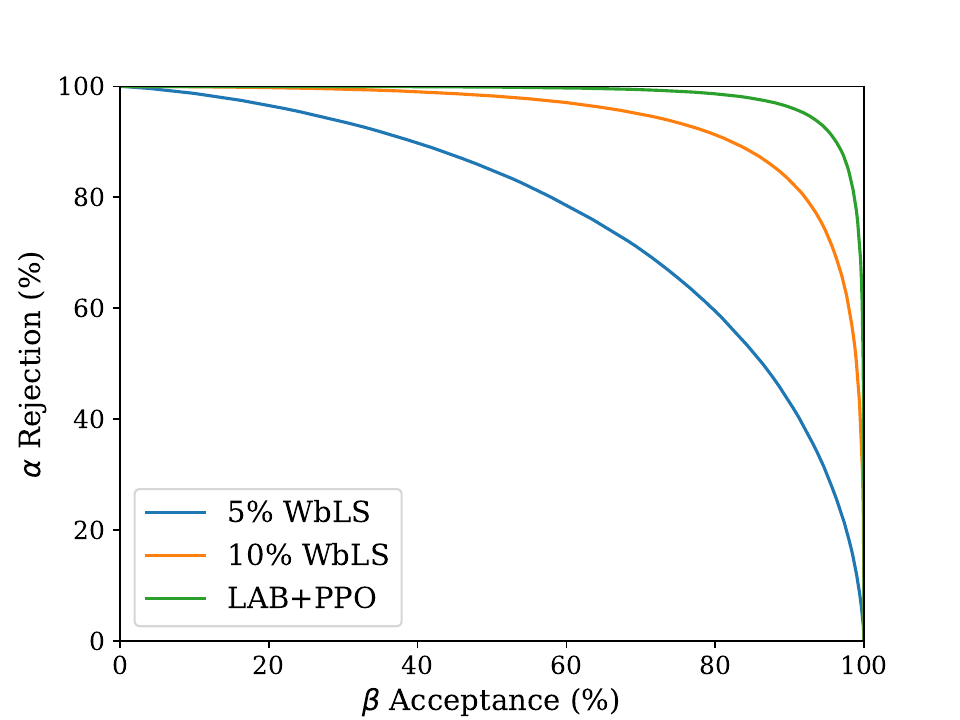}
    \includegraphics[width=0.65\columnwidth]{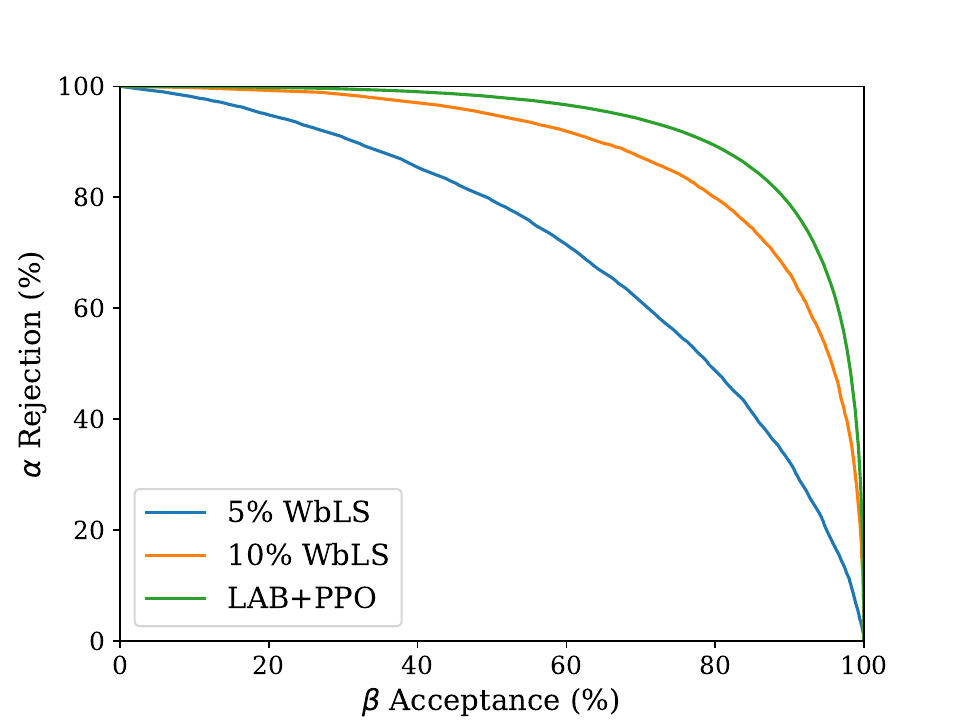}
    \includegraphics[width=0.65\columnwidth]{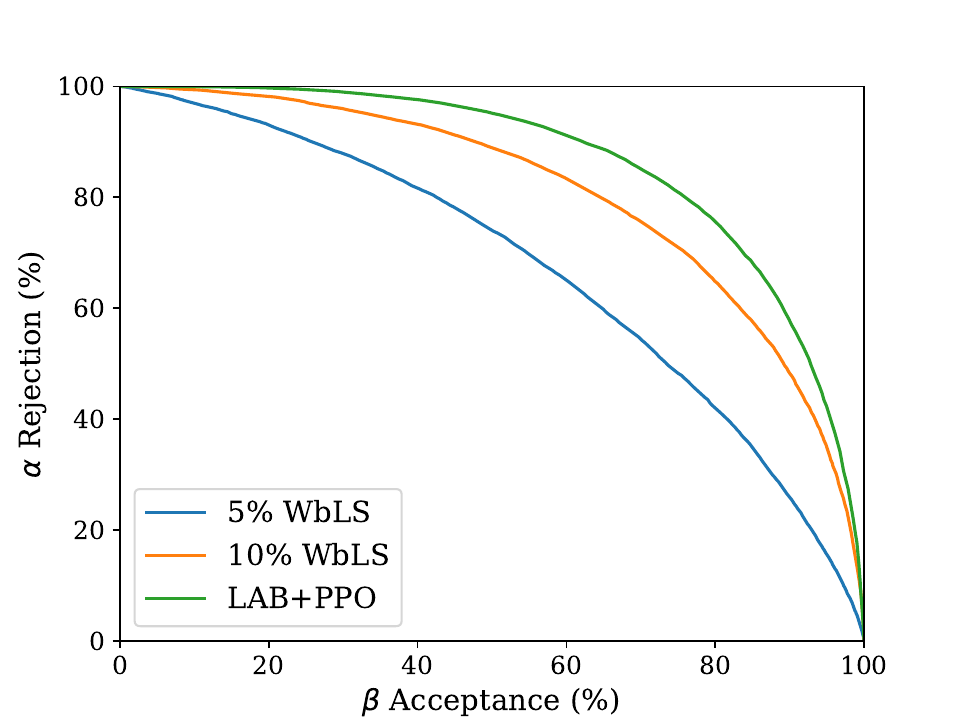}
    \caption{Comparisons of the simultaneously achievable $\beta$ acceptance and $\alpha$ rejection, using all detected PEs, for 5\% \wbls{}, 10\% \wbls{}, and LAB+PPO for the \SI{4}{\tonne{}} (left), \SI{1}{\kilo\tonne{}} (center) and \SI{100}{\kilo\tonne{}} (right) detectors. All events are simulated at the center of the detector, uniformly in direction. The $\alpha$ particles are produced from $^{210}$Po decays and the $\beta$ particles are simulated with energies from Table \ref{tab:betaen}.}
    \label{fig:pidroc}
\end{figure*}

    Higher light yield materials provide better PID performance, as a result of the classifier variance decreasing in accordance with the central limit theorem. As evidenced in Fig.~\ref{fig:rejsummaryplot}, scattering and absorption of photons as they propagate through ever-larger detectors can have a substantial impact, as the smaller, lower photocoverage \SI{4}{\tonne{}} detector outperforms the larger, higher photocoverage \SI{}{~\kilo\tonne{}}-scale detectors. This is in accordance with the washing out of features in time-residual PDFs for larger detectors (Fig.~\ref{fig:pidpdfsmultikt}) due to smearing from absorption/reemission and optical scattering. Additionally, we find that the particle identification performance ranking of the various detector configurations is robust to the choice of figure of merit used to generate the cuts. Table~\ref{table:rejresults} shows the performance for 90\%, 99\%, and 99.9\% $\alpha$ rejection to directly compare to existing experiments, such as Borexino \cite{Basilico:2023xui} and SNO+ \cite{SNO:2020fhu}. We also show in Fig.~\ref{fig:pidroc} the simultaneously achievable $\beta$ acceptance and $\alpha$ rejection for the examined scenarios.

A consequence of the short time window used to measure the emission timing in Sect.~\ref{sec:timing} is in an underestimation of the amount of light emitted at late times, beyond 60~ns. This yields a conservative estimation of the expected PID performance for the LAB+PPO, and to a lesser extent, the WbLS. We repeated the PID analysis using the published LAB+PPO emission timing measured by the SNO+ detector~\cite{SNO:2020fhu} for all three detector configurations. These measurements use a longer analysis window and fit using a four-decay exponential model. We find that using the SNO+ emission timing results in 100\% separation of the $\alpha$ and $\beta$ particles in all three detector configurations.

\begin{table}[t!]
\centering
\caption{$\beta$ acceptance in \SI{4}{\tonne{}}, \SI{1}{\kilo\tonne{}} and \SI{100}{\kilo\tonne{}} detectors for various materials, for cut values that yield the stated $\alpha$ rejection. Notably, as discussed in Section \ref{sec:pidres}, using the LAB+PPO emission timing measured in~\cite{SNO:2020fhu} results in 100\% separation for all three detector configurations.}
\begin{tabular}{c|c|c|c|c}
\hline\hline
    Det.     & Material   &  Threshold  & Cut Value &   $\beta$ Acc. [\%] \\
\hline
 \SI{4}{~\tonne{}}        & 5\% \wbls{}  & 90.0\%      &      -0.016 &                      39.3   \\
 \SI{4}{~\tonne{}}        & 5\% \wbls{}  & 99.0\%      &      -0.034 &                       8.5   \\
 \SI{4}{~\tonne{}}        & 5\% \wbls{}  & 99.9\%      &      -0.049 &                       1.3   \\
 \SI{4}{~\tonne{}}        & 10\% \wbls{} & 90.0\%      &      -0.005 &                      82.3   \\
 \SI{4}{~\tonne{}}        & 10\% \wbls{} & 99.0\%      &      -0.027 &                      40.9   \\
 \SI{4}{~\tonne{}}        & 10\% \wbls{} & 99.9\%      &      -0.043 &                      13.4   \\
 \SI{4}{~\tonne{}}        & LAB+PPO      & 90.0\%      &       0.001 &                      96.3   \\
 \SI{4}{~\tonne{}}        & LAB+PPO      & 99.0\%      &      -0.002 &                      76.8   \\
 \SI{4}{~\tonne{}}        & LAB+PPO      & 99.9\%      &      -0.004 &                      48.1   \\
 \hline
 \SI{1}{~\kilo\tonne{}}   & 5\% \wbls{}  & 90.0\%      &      -0.012 &                      31.8   \\
 \SI{1}{~\kilo\tonne{}}   & 5\% \wbls{}  & 99.0\%      &      -0.027 &                       5.9   \\
 \SI{1}{~\kilo\tonne{}}   & 5\% \wbls{}  & 99.9\%      &      -0.039 &                       0.6   \\
 \SI{1}{~\kilo\tonne{}}   & 10\% \wbls{} & 90.0\%      &      -0.007 &                      64.6   \\
 \SI{1}{~\kilo\tonne{}}   & 10\% \wbls{} & 99.0\%      &      -0.022 &                      26.1   \\
 \SI{1}{~\kilo\tonne{}}   & 10\% \wbls{} & 99.9\%      &      -0.035 &                       5.6   \\
 \SI{1}{~\kilo\tonne{}}   & LAB+PPO      & 90.0\%      &      -0.000 &                      79.1   \\
 \SI{1}{~\kilo\tonne{}}   & LAB+PPO      & 99.0\%      &      -0.003 &                      40.6   \\
 \SI{1}{~\kilo\tonne{}}   & LAB+PPO      & 99.9\%      &      -0.004 &                      16.0   \\
 \hline
 \SI{100}{~\kilo\tonne{}} & 5\% \wbls{}  & 90.0\%      &      -0.008 &                      25.7   \\
 \SI{100}{~\kilo\tonne{}} & 5\% \wbls{}  & 99.0\%      &      -0.018 &                       4.0   \\
 \SI{100}{~\kilo\tonne{}} & 5\% \wbls{}  & 99.9\%      &      -0.024 &                       0.7   \\
 \SI{100}{~\kilo\tonne{}} & 10\% \wbls{} & 90.0\%      &      -0.006 &                      47.7   \\
 \SI{100}{~\kilo\tonne{}} & 10\% \wbls{} & 99.0\%      &      -0.015 &                      13.6   \\
 \SI{100}{~\kilo\tonne{}} & 10\% \wbls{} & 99.9\%      &      -0.022 &                       3.2   \\
 \SI{100}{~\kilo\tonne{}} & LAB+PPO      & 90.0\%      &      -0.001 &                      61.9   \\
 \SI{100}{~\kilo\tonne{}} & LAB+PPO      & 99.0\%      &      -0.002 &                      29.7   \\
 \SI{100}{~\kilo\tonne{}} & LAB+PPO      & 99.9\%      &      -0.004 &                      13.8   \\
\hline
\end{tabular}
\label{table:rejresults}
\end{table}

\subsection{Discussion}\label{sec:pid_discussion}

    The results presented in Sect.~\ref{sec:pidres} use only timing information, considered over a fixed analysis window. A pertinent question is the role of the Cherenkov light produced by electrons, as its higher proportion in \wbls{} samples may be expected to contribute to timing-based PID beyond that afforded by differences in scintillation emission time profiles. By ignoring Cherenkov photons in the PID analysis of a 5\% \wbls{} filled EOS-like detector, we surprisingly find that the presence of Cherenkov light degrades $\alpha$ rejection at the level of 1.5\%. This is due to the Cherenkov component competing with the scintillation light rise time, which is measured to be larger for $\beta$s than for $\alpha$s. Thus, at \isotope{Po}{210} energies, the small amount of Cherenkov light emitted for $\beta$s causes the effective rise-time to look more similar to $\alpha$s, degrading the classification power.  

    At higher light levels, produced by higher energy $\beta$s, the larger Cherenkov contribution is sufficient to enact a genuine shape difference in time profiles, and enhances PID performance. For example, at the light levels associated with \isotope{Po}{212} $\alpha$ decays, the behavior is reversed, and the Cherenkov light provides a roughly $\sim 1.5\%$ increase in $\alpha$ rejection, absolute. Of course, it should be noted that the measured scintillation rise times may be subject to systematic uncertainties associated with choices in system response modeling, potentially affecting the competition between the Cherenkov component and the scintillation rise times discussed here.

    It should be noted that hybrid detectors which leverage dedicated techniques to identify, i.e. ``tag,'' Cherenkov photons may achieve PID performance beyond that available via the simple likelihood-ratio statistic employed here, owing to the inclusion of other observables. Examples of such observables are angular information and wavelength, employed via the observation of anisotropic photon collection, after performing vertex and direction reconstruction, and spectral photon sorting, respectively. The timing, topological, and spectral information could, in principle, be combined in a joint vertex-direction-PID fit, which inherently determines a particle's identity based on the interaction geometry. These extensions are promising, but their technical exploration is beyond the scope of this work. Lastly, it is to be expected that the PID would improve in larger detectors, compared to what is shown here, because of the ability to leverage differences in the time profile over a longer time window.

%% file: conclusion.tex
\section{Conclusion}\label{sec:conclusion}
This work presents the first characterization of the response of \wbls{} samples to $\alpha$ radiation. Using a simulation model, the total amount of scintillation light produced by a \SI{4.8}{~\MeV} $\alpha$ particle in 1\%, 5\%, and 10\% \wbls{} is measured, and the results are interpreted in the context of ionization quenching using Birks' law. The scintillation emission time profiles of 5\% and 10\% \wbls{} are determined using an analytic model, which includes detailed modeling of the system response. These material properties are used as input parameters for MC simulations used to make predictions for real detectors, ranging in scale from from few-ton to $\SI{100}{~\kilo\tonne}$.

Greater PID performance is observed in higher light yield materials and smaller detector sizes, despite lower photocoverage, owing to the increased importance of optical effects at play when photons traverse larger distances. These results utilize only timing information over a limited analysis window, and thus provide a baseline performance which may be improved upon by dedicated techniques leveraging the topological and spectral features of Cherenkov light and by utilizing additional scintillation light across a broader window.

Neutrino experiments, such as Borexino, have found that $\alpha$ rejection is critical in studying low energy CNO solar neutrinos \cite{Borexino:2013zhu}. Thus, future experiments using \wbls{} will likely need to utilize $\alpha$/$\beta$ discrimination to improve solar neutrino sensitivity. The results presented in this manuscript allow for more realistic solar neutrino studies in \wbls{}, that build on work detailed in Ref. \cite{Bonventre:2018hyd}. This will allow a better understanding of the solar neutrino physics reach of hybrid detectors, although such advanced detectors will likely use additional technology and analysis techniques to improve $\alpha$/$\beta$ PID beyond the timing-only PID studied in this manuscript.
\vfill

%% file: acknowledgements.tex
\section*{Acknowledgements}

 Work conducted at Lawrence Berkeley National Laboratory was performed under the auspices of the U.S. Department of Energy under Contract DE-AC02-05CH11231. The work conducted at Brookhaven National Laboratory was supported by the U.S. Department of Energy under contract DE-AC02-98CH10886. EJC was funded by the Consortium for Monitoring, Technology, and Verification under Department of Energy National Nuclear Security Administration award number DE-NA0003920. The project was funded by the U.S. Department of Energy, National Nuclear Security Administration, Office of Defense Nuclear Nonproliferation Research and Development (DNN R\&D). This material is based upon work supported by the U.S. Department of Energy, Office of Science, Office of High Energy Physics, under Award Number DE-SC0018974.